\documentclass[iop, onecolumn, apj, tighten, numberedappendix]{emulateapj}
\slugcomment{{To appear in the Astrophysical Journal.}}

\usepackage{amsmath,amsthm}
\usepackage{amssymb}

\usepackage{mathtools}
\usepackage{empheq}
\usepackage{bbm}
\usepackage{bm}

\usepackage{epstopdf}

\usepackage{color} 

\newcommand{\myemail}{vboening@kis.uni-freiburg.de}

\newcommand\ii{{\rm i}}
\newcommand\id{{\mathrm d}}

\newcommand\cO{{\mathcal O}}
\newcommand\cL{{\mathcal L}}
\newcommand\br{{\mathbf r}}
\newcommand\brobs{{{\bf r}_\text{obs}}}
\newcommand\bl{{\bf l}}

\newcommand\Power{{\mathcal{P}}}

\newcommand\bnabla{{\bm \nabla }}
\newcommand\bcdot{{\, \bm \cdot \,}}
\renewcommand\bv{{\bf v}}
\newcommand\bS{{\bf S}}
\newcommand\bxi{{\bm \xi}}

\newcommand\bG{{\rm \bf G}}
\newcommand\cG{{\mathcal G}}

\newcommand\cF{{\mathcal F}}
\newcommand\cW{{\mathcal W}}

\newcommand\bK{{ \bf K}}
\newcommand\unit{{\bf\hat{e}}}
\newcommand\plainunit{{\hat{e}}}

\newcommand\Cref{C_{\text{ref}}}
\newcommand\cref{\Cref}
\newcommand\dotCref{\dot C_{\text{ref}}}

\newcommand\taudiff{\tau_{\text{diff}}}

\newcommand\tauomega{{\delta \tau_{\text{diff}}^\Omega}}
\newcommand\taukernel{{\delta \tau_{\text{diff}}^K}}
\newcommand\Wdiff{{W_{\text{diff}}}}
\newcommand\Wdiffstar{{W^*_\text{diff}}}

\newcommand\comega{{C_\Omega}}

\newcommand\COmega{\comega}

\newcommand\bZ{{\bf Z}}

\newcommand\robs{{r_{\text{obs}}}}
\newcommand\EE{{\mathbb{E}}}

\newcommand\bcurlyC{{\bm{\mathcal{C}}}}

\shorttitle{Spherical Born Kernels for Flows}

\shortauthors{B\"oning et al.}

\begin{document}


\title{Sensitivity Kernels for Flows in Time-Distance Helioseismology: \\
 Extension to Spherical Geometry}

\author{Vincent~G.~A.~B\"oning, Markus~Roth, and Wolfgang~Zima}
\affil{Kiepenheuer-Institut f\"ur Sonnenphysik, 79104 Freiburg, Germany}
\email{\myemail}
\author{Aaron~C.~Birch}
\affil{Max-Planck-Institut f\"ur Sonnensystemforschung, 37077 G\"ottingen, Germany}
\and
\author{Laurent~Gizon}
\affil{Max-Planck-Institut f\"ur Sonnensystemforschung, 37077 G\"ottingen, Germany}
\affil{Institut f\"ur Astrophysik, Georg-August-Universit\"at G\"ottingen, 37077 G\"ottingen, Germany}


\begin{abstract}

We extend an existing Born approximation method for calculating the linear sensitivity of helioseismic travel times to flows from Cartesian to spherical geometry. This development is necessary for using the Born approximation for inferring large-scale flows in the deep solar interior. 
In a first sanity check, we compare two $f-$mode kernels from our spherical method and from an existing Cartesian method. The horizontal and total integrals agree to within 0.3~\%. 
As a second consistency test, we consider a uniformly rotating Sun and a travel distance of 42 degrees. The analytical travel-time difference agrees with the forward-modelled travel-time difference to within 2~\%. 
In addition, we evaluate the impact of different choices of filter functions on the kernels for a meridional travel distance of 42 degrees. 
For all filters, the sensitivity is found to be distributed over a large fraction of the convection zone. We show that the kernels depend on the filter function employed in the data analysis process. 
If modes of higher harmonic degree ($90\lesssim l \lesssim 170$) are permitted, a noisy pattern of a spatial scale corresponding to $l\approx 260$ appears near the surface. 
When mainly low-degree modes are used ($l\lesssim70$), the sensitivity is concentrated in the deepest regions and it visually resembles a ray-path-like structure.
Among the different low-degree filters used, we find the kernel for phase-speed filtered measurements to be best localized in depth.

\end{abstract}




\section{INTRODUCTION}
\label{intro}

\notetoeditor{We would prefer a single column format due to the large widths of the equations.}Accurate measurements of flows in the deep solar interior are important for understanding meridional circulation, differential rotation, and giant convection cells. 
Time-distance helioseismology \citep{Duvall1993} is one of several methods capable of measuring flows in the solar interior. Inferences of deep flows have been obtained using local and global helioseismology (e.g., \citealp{Giles1997Nature}, \citealp{GilesPhD}, \citealp{Beck2002}, \citealp{Basu2003}, \citealp{Howe2009}, \citealp{Schad2012}, \citealp{Woodard2013}, \citealp{Schad2013}).

For deep flow measurements using time-distance helioseismology, large travel distances of 40 degrees in heliocentric angle and beyond are used (for recent results, see \citealp{Zhao2013}, \citealp{KholikovHill2014}, \citealp{Kholikov2014}, \citealp{Jackiewicz2015}, \citealp{Liang2015SurfaceMagnOnMeridFlow}, \citealp{Liang2015ProbingMagneticFields}, \citealp{Rajaguru2015}). The necessary forward modelling in spherical geometry of the effect of a flow field on the measured travel times has, so far, always been done using the ray approximation (\citealp{Kosovichev1996}; \citealp{KosovichevDuvall1997}). It is an infinite frequency approximation in which travel times are only sensitive to flows along a ray path. For smaller travel distances, where the geometry can be considered Cartesian, both ray and Born approximations have been used for the forward modelling (e.g., \citealp{Zhao2012}, \citealp{Svanda2011}).

In the context of helioseismology, the Born approximation was used by \citet{Birch2000} to obtain the sensitivity of travel-time measurements to perturbations in sound speed. It assumes that the sensitivity of the travel time of a wave packet to flows is caused by a scattering process which can take place at any location inside the Sun. This scattering process is modelled at first order and takes a range of finite mode frequencies into account \citep{GB2002,Sakurai2011}. A general recipe for calculating the sensitivity of helioseismic travel times to any perturbative quantity using the Born approximation was developed by \citet{GB2002}. Sensitivity functions (also known as kernels) for flows were obtained with this method by, e.g., \cite{Gizon2000}, \citet{BG2007}, \citet{Birch2007}, \citet{Jackiewicz2007a}, and \citet{Burston2015}, where the underlying geometry was assumed to be Cartesian.

The accuracy of the Born and ray approximations has been studied by, e.g., \citet{Bogdan1997}, \citet{Birch2004}, \citet{Couvidat2006}, and \citet{BG2007}. The ray approximation is expected to be valid when the length scale of variations of the flow is larger than the width of the first Fresnel zone (e.g. \citealp{Hung2000}, \citealp{Birch2001}). In a uniform medium, the width of the first Fresnel zone is $L\approx\, \sqrt{\lambda \Delta}$ (e.g., \citealp{Gizon2006}), where $\lambda$ is the wavelength and $\Delta$ is the travel distance.  Using this simple approximation, we estimate $L \approx 200\,\rm{Mm}$ for a time-distance measurement probing the base of the convection zone.  It is not known if flows deep in the solar interior are smooth on this length scale.  It is thus important to carry out forward modelling (and eventually inversions) using a finite wavelength approximation in place of the ray approximation  (e.g., \citealp{Zhao2013}, \citealp{Jackiewicz2015}, \citealp{Rajaguru2015}).

First attempts to calculate Born approximation sensitivity functions for flows in spherical geometry have been made by \citet{RGB2006}, where preliminary results were presented. In this paper, we extend the method of \cite{BG2007} for calculating the linear sensitivity of helioseismic travel times to flows with the Born approximation from Cartesian to spherical geometry. Similar to the work of \cite{BG2007}, our procedure is largely based on \cite{GB2002}. 
We first introduce some notation necessary to describe the measurement process in Section \ref{secobservations} and present the derivation of the zero-order solution to the model in Section \ref{seczeroorder}. In Section \ref{seckernelformula}, we derive the first-order solution and find a spherically geometric kernel formula, the numerical implementation of which is discussed in Section \ref{secnumerics}. 
Section~\ref{secsanity} includes a comparison of results from our code to the Cartesian model of \cite{BG2007}, which serves as a first sanity check of our computations. In Section \ref{secdeepkernels}, we present some example kernels for large travel distances which could be used for measuring, e.g., the deep meridional flow. The results are tested for self-consistency and used to evaluate different filtering schemes. Finally, we summarize and discuss our results in Section~\ref{secdiscussion}.

\section{MODELLING TRAVEL-TIME MEASUREMENTS}
\label{secobservations}

The general procedure in time-distance helioseismology \citep{Duvall1993} is to measure travel times of waves travelling between two different locations on the Sun and to infer solar interior properties from shifts in these travel times, $\delta\tau$, with respect to a reference model. The measured travel-time shifts, $\delta \tau$, which are modelled as stochastic variables, can be split into their expectation value, $\EE[\delta \tau]$, and in a noise component, $\epsilon$ \citep[see][]{Couvidat2005},
\begin{equation}
\delta \tau= \EE[\delta \tau] + \epsilon.
\label{eqttobs}
\end{equation}
Whenever the travel-time shifts can be assumed to be caused by weak solar interior flows, we can assume a linear relationship between a flow field $\bv(\br)$ in the solar interior and the expected travel-time shift,
\begin{equation}
\EE[\delta \tau] = \int_\Sun \bK(\br) \bcdot \bv(\br) \, \id^3 \br.
\label{eqkernelgoal}
\end{equation}
It is the goal of this paper is to derive such a relationship using the Born approximation \citep{GB2002} and thus to derive the travel-time sensitivity function $\bK(\br)$. Equation~\eqref{eqkernelgoal} may then be used to perform inversions of the solar interior flow field using equation~\eqref{eqttobs}.

As this goal has been achieved in Cartesian geometry by \citet{BG2007}, we extend their work to spherical geometry and largely follow their development including the use of the theoretical framework developed by \cite{GB2002}. In our model, we thus incorporate the whole measurement process as well as the first-order perturbation to the solar interior wave field caused by any weak and steady flow in the solar interior.

\subsection{Modelling the Doppler Signal}

Travel times are obtained from an observational Doppler signal $\tilde\Phi(\br,t)$ at locations $\brobs=(r_{\text{obs}},\theta,\phi)$ at time $t$. The arguments $r_{\text{obs}}$, $\theta$, and $\phi$ denote distance to the center of the Sun, colatitude, and longitude, respectively, of the location on the solar surface, where the signal is originating. We use a tilde for unfiltered variables and bold symbols for vector quantities throughout this paper. The Doppler signal can be modelled as a line-of-sight projected velocity of the surface oscillation,
\begin{align}
        \tilde \Phi(\brobs,t) &= \hat\bl(\brobs) \bcdot \dot \bxi(\brobs,t), \label{equnfilteredsignal}
\end{align}
where $\bxi(\brobs,t)$ is the oscillatory displacement vector and $\hat\bl(\brobs)$ is unit vector in the line-of-sight direction at the same location. The time derivative of a variable is denoted by a dot over the respective variable.

In this paper, we assume the line-of-sight operator to be radial, $\hat\bl \equiv \unit^{(r)}$, where $\unit^{(k)}$ is the unit vector in direction $k=r,\theta,\phi$ with components $\plainunit^{(k)}_j=\delta_{jk}$ at location $\br$ and $\delta_{jk}$ is the Kronecker delta. This assumption simplifies our computations although it is incorrect for observations from a fixed point-of-view especially in the case of large travel distances. We choose this rather simple model since it is comparable to existing results of kernel calculations \citep{BG2007} in terms of a radial line-of-sight projection.

The unfiltered signal is thus assumed to be $\tilde \Phi(\brobs,t) = \dot \xi_r(\brobs,t)$ or $\tilde \Phi(\brobs,\omega) = -\ii \omega \xi_r(\brobs,\omega)$, where, in the time and frequency domains, we use the Fourier transform convention of \cite{GB2002} repeated in Appendix \ref{appendixfourier}. Note that we indicate the Fourier transform of a function by the use of the Fourier space variable, in this case the angular frequency $\omega$. Following \cite{GB2002}, for a given observational time duration $T$, we assume all Fourier transformed time-dependent quantities to be truncated to zero for $|t|\ge T/2$. The time interval $T$ is assumed to be sufficiently large that the effect of the truncation can be neglected and that the expectation values considered in the modelling process can attain the appropriate limits, see \citet{GB2002} and \citet{Fournier2014}.

\subsection{Filtering and Power Spectra}

The signal $\tilde\Phi$ is filtered in the data analysis process in order to, e.g., select specific waves which travel a similar distance and to increase the signal-to-noise ratio for specific measurements, see, e.g., \citet{Couvidat2006phase1} and \cite{Zharkov2006}. This is done using the unfiltered spherical harmonic time series, which is obtained according to 
\begin{equation}
\tilde a_{lm}(t) = \int_{S^2} Y_{lm}^*(\Omega) \tilde \Phi(\robs,\Omega,t) \, \id \Omega. \label{eqSHT}
\end{equation}
Here, the integration domain is the unit sphere $S^2$ parametrized by $\Omega=(\theta,\phi)$ and the $Y_{lm}(\theta,\phi)$ are spherical harmonics of harmonic degree $l$ and azimuthal order $m$. In equation~\eqref{eqSHT}, we make the simplifying assumption that the data are available for all spatial positions and that they are observed everywhere at the same geometrical height $\robs$. The filter is applied by multiplying a filter function $f(l,\omega)\geq0$ in the $(l,\omega)$ domain (see also \citealp{Kholikov2014}),
\begin{align}
        a_{lm}(\omega) &= f(l,\omega) \tilde a_{lm}(\omega) \label{eqalmfiltered2}.
\end{align}
The filtered signal is reconstructed from an inverse Fourier transform and an inverse spherical harmonic transform,
\begin{align}
        \Phi(\theta,\phi,t) &= \sum_{l,m} a_{lm}(t) Y_{lm}(\phi,\theta).
        \label{eqdatafiltered2}
\end{align}

Additionally, a filtered $m$-summed power spectrum can be obtained in the $(l,\omega)$ domain via
\begin{align}
   \Power(l,\omega) &= \frac{2\pi}{T}\sum_{m=-l}^l |a_{lm}(\omega)|^2 = f(l,\omega)^2 \;\frac{2\pi}{T} \sum_{m=-l}^l |\tilde a_{lm}(\omega)|^2 = f(l,\omega)^2 \; \tilde \Power(l,\omega),
	\label{eqpowerfiltered}
\end{align}
where $\tilde \Power(l,\omega)$ is the unfiltered power spectrum.

\subsection{Cross-Covariance Functions}

From the filtered signal $\Phi$ at two observation points $\br_j=(r_j,\theta_j,\phi_j)=(r_j,\Omega_j),\, j=1,2$, cross-covariance functions are obtained from equation~(3) in \citet{GB2002} or equivalently in frequency space,
\begin{equation}
        C(\br_1,\br_2,\omega) = \frac{2 \pi}{T} \Phi^*(\br_1,\omega)  \Phi(\br_2,\omega). \label{eqconvcorrw}
\end{equation}

In practice, before computing cross-covariance functions, the signal is usually averaged in space (e.g., \citealp{GB2005}). The cross-covariance function of spatially averaged signals can in turn be written as an average over point-to-point cross-covariance functions. Therefore, we only consider point-to-point measurements in this paper.

\subsection{Travel-Time Fitting}

In order to obtain travel times from a cross-covariance function, we use the travel-time definition of \citet[eq. B5]{GB2004}, see also \citet[eqs. A6 and A8]{GB2002},
\begin{align}
        \delta \tau_a(\br_1,\br_2) &= \int_{-\infty}^\infty W_a(\Delta_{1,2},t) \left[ C(\br_1,\br_2,t) - \Cref(\Delta_{1,2},t)\right] \; \id t, \label{eqtau}
\end{align}
where $a \in \{+,-,\text{diff}\}$. The weight functions $W_\pm$ are defined as				
	\begin{align}			
	     W_\pm(\Delta,t) &= \frac{ \mp h(\pm t) \dotCref(\Delta,t)}{\int_{-\infty}^\infty h(\pm t') [\dotCref(\Delta,t')]^2 \, \id t'} \label{eqW}.
\end{align}
The symbol $\Delta_{1,2}=\Delta(\br_1,\br_2)$ denotes the angular distance between the two observation points $\br_1$ and $\br_2$. The function $\Cref(\Delta,t)$ is a direction-independent reference cross-covariance function. It is derived from our model in the following section. The window function $h$ is used to select a positive time part of the cross-covariance function, e.g. an interval around the first bounce travel time (see \citealp{GB2002} or \citealp{BG2007}). For measuring flows, we consider $\delta \taudiff=\delta \tau_+ - \delta\tau_-$ and consequently $\Wdiff = W_+ - W_-$ in this paper (see also \citealp{GB2002}).

\section{EXPRESSIONS FOR ZERO-ORDER QUANTITIES}

\label{seczeroorder}

Before we can model the effect of a flow field on the solar interior wave field and thus on the travel times, we have to solve for the wave field in zero order, that is in the absence of flows. Without a flow field, the expected travel-time shift is zero, $\EE[\delta\taudiff]=0$. When flows are present, their first-order effect on the wave field, on the expectation value of the cross-covariance function, and on the expectation value of the travel-time shift can then be modelled, see Section \ref{seckernelformula}.

\subsection{Zero-Order Wave Equation}
\label{seczeroorderequation}
We assume a spherically symmetric, non-rotating Sun without interior flows as given by solar model~S \citep{JCD1996}. The wave equation for stochastically excited and damped solar acoustic oscillations thus reads (see also \citealp{GB2002} and \citealp{BG2007})
\begin{align}
\cL[\bxi] &= \bS \label{eqzeroorder},
\end{align}
where
\begin{equation}
\cL [\bxi(\br,\omega)]  \equiv \rho_0 (-\omega^2 -2\ii\omega\Gamma + \cW )[\bxi(\br,\omega)]  \label{eqLfreqspace} 
\end{equation}
and where $\bS$ models the stochastic sources. Here, $\bxi(\br,t)$ is the oscillatory displacement vector in the solar interior at location $\br$ at time $t$, $\cW$ is the linear wave operator, $\Gamma$ describes the damping, and $\rho_0$ denotes density.

The form of the wave operator $\cW$ is described by \citet[$\cF$ in eq. 3.245]{Aerts2010}. 
The eigenfunctions of adiabatic solar oscillations without damping, $\bxi^{lmn}(\br)$, which correspond to standing waves in the solar interior, satisfy
\begin{equation}
\cW [\bxi^{lmn}(\br)] = \omega_{ln} ^2 \; \bxi^{lmn}(\br),
\label{eqeigenfunctions2}
\end{equation}
where $l$ is the harmonic degree, $m$ the azimuthal order, and $n$ the radial order of the mode with unperturbed eigenfrequency $\omega_{ln}$. For the spherical kernels in this paper, eigenfunctions and eigenfrequencies were calculated with solar model~S \citep{JCD1996} and ADIPLS \citep{JCDadipls}. We assumed the standard ADIPLS surface boundary condition of a vanishing Lagrangian pressure perturbation at the surface during oscillations (see \citealp{JCDadipls}).

\citet{BirchKo2004} note that it is important that mode frequencies, damping rates and amplitudes in the model power spectrum have to match those from observations. In order to incorporate the damping rates in an easy and accurate manner into our model, we assume the damping operator $\Gamma$ to have the same eigenmodes as the wave operator,
\begin{equation}
\Gamma \bxi^{lmn} = \gamma_{ln} \bxi^{lmn}. \label{eqdamping}
\end{equation}
In practice, damping rates $\gamma_{ln}$ are used from observations, see Sections \ref{secsanity} and \ref{secdeepkernels}.

We apply a source model similar to \cite{GB2002}, \cite{BirchKo2004}, and \cite{BG2007} and assume that the source function $\bS(\br,t)$ can be modelled by a stationary stochastic process with source covariance
\begin{align}
M_{ij}(\br',\br'';\omega) =\frac{2\pi}{T} \, \EE [S_i^*(\br',\omega)S_j(\br'',\omega)]
\stackrel{(*)}{\cong} M(\omega) \, \delta(\br'-\br'') \delta_{i,r} \delta_{j,r}  \delta(|\br'|-r_s) \frac{1}{r_s^2},
\label{eqsourcecovariance}
\end{align}
where $M(\omega)$ is the spectral density function of the sources chosen according to \cite{BirchKo2004}. Sources at two different locations $\br'\neq\br''$ and two different directions $i\neq j$ ($i,j\in\{r,\theta,\phi\}$) of a source are assumed to be uncorrelated. Furthermore, the relation (*) is valid in the limit of a large time interval $T$, see \citet[eq. 4.7.5.]{Priestley1980}. 
As the multiplication of \eqref{eqsourcecovariance} by a constant factor does not alter the kernel result, we introduced a factor of $1/r_s^2$ in equation~\eqref{eqsourcecovariance} for simplicity in the following computations, where $r_s$ is the distance of the sources to the solar center.

\subsection{Zero-Order Solution via Green's Functions}
\label{seczeroordersolution}

The wave field which solves the zero-order problem can be written in terms of eigenmodes. The eigenfunctions $\bxi^{lmn}(\br)$ of solar oscillation from equation~\eqref{eqeigenfunctions2} can be written (see \citealp{Aerts2010})
\begin{align}
\bxi^{lmn}(\br) &=  \left[ R_{ln}(r) \unit^{(r)} + \frac{H_{ln}(r)}{\sqrt{l(l+1)}} \left( \unit^{(\theta)} \partial_\theta + \frac{\unit^{(\phi)}}{\sin\theta} \partial_\phi\right) \right]  {Y}_{lm}(\theta,\phi)  \label{eqxiRH} \\
&\equiv \sum \limits_{k=r,\theta,\phi} \unit^{(k)}(\br) \cO_k^{ln}(r) \left[ {Y}_{lm}(\theta,\phi) \right], \label{eqxiRH2}
\end{align}
where $R_{ln}$ and $H_{ln}$ are radial and horizontal eigenfunctions, $\partial_\theta$ and $\partial_\phi$ are the derivatives with respect to $\theta$ and $\phi$, respectively, and we use $\cO_k^{ln}$ to abbreviate the eigenfunction-dependent differential operators which act on the spherical harmonics in equation~\eqref{eqxiRH}. We use the normalization condition (e.g., \citealp{Birch2000})
\begin{equation}
        \int_{\sun}
				\rho_0(r)  \bxi^{lmn*}(\br) \bcdot \bxi^{l'm'n'}(\br)\, \id^3\br =  \delta_{l,l'}\delta_{m,m'}\delta_{n,n'} \label{eqorthonormal}.
\end{equation}

The solution is obtained with the help of Green's functions $\bG^k$ (see \citealp{BG2007}) which satisfy
\begin{equation}
\cL \left[ \bG^k(\br|\br',\omega) \right] =  \unit^{(k)}(\br') \delta(\br-\br'), \; \; (k=r,\theta, \phi). 
\label{eqgreens}
\end{equation}
Equation~\eqref{eqgreens} is written in frequency space and may differ by a factor of $2\pi$ from what one may expect, see equation~\eqref{eqgreenstimedomaincondition}. From equations \eqref{eqgreens} and \eqref{eqzeroorder}, we obtain for the zero order solution $\bxi$,
\begin{equation}
\xi_j(\br,\omega) =  \int_\sun   G^k_j(\br|\br',\omega) S_k(\br',\omega) \, \id^3 \br' , \; \; (j=r,\theta, \phi).
\label{eqansatz}
\end{equation}
Note that throughout this work, we employ the convention that repeated indexes are summed over, in this case over $k=r,\theta,\phi$.
See Appendix \ref{appendixgreens} for a detailed derivation of the expression for the Green's functions from equation~\eqref{eqgreens}, from which we can obtain an expression for the filtered Green's function,
\begin{align}
        {\cG}^k(\brobs|\br',\omega) 
				= \sum_{lmn} f(l,\omega)  \frac{  \xi_k^{lmn*} (\br') \xi_r^{lmn}(\brobs) }{\sigma^2_{ln}-\omega^2 }, \label{eqfilteredgreens}
\end{align}
where we took advantage of our simplifying assumptions on the filtering procedure and used equation~\eqref{eqgreensforfiltering}. The filtered Doppler signal due to solar oscillations can thus be modelled by
\begin{align}
        \Phi(\brobs,\omega)  
				= -\ii\omega  \int_\sun { \cG^k(\brobs|\br',\omega)}  S_k(\br',\omega) \, \id^3 \br'. \label{eqdopplerfilteredgreens} 
\end{align}

\subsection{Zero-Order Power Spectrum}
\label{seczeroorderpower}

An expression for the zero-order power spectrum $\Power_0(l,\omega)$ can be derived as an expectation value of the power spectrum when no flows are present in the model. It is obtained from the zero-order Doppler signal, see Appendix \ref{appendixpower}, as
\begin{align}
         \Power_0(l,\omega) &=  \omega^2 M(\omega)  f(l,\omega)^2 (2l+1)\sum_{nn'} \frac{ R_{ln}(\robs) R_{ln'}(\robs) R_{ln}(r_s)R_{ln'}(r_s) }{(\sigma^{2*}_{ln}-\omega^2)(\sigma^{2}_{ln'}-\omega^2)} \label{eq0power}.
\end{align}

In order to show that equation~\eqref{eq0power} reproduces power spectra from observations, we follow \citet{BirchKo2004} for an approximation to this expression near a resonance peak. First, we have, 
\begin{equation}
	\sigma^{2}_{ln}-\omega^2 = - (\omega-\omega_{ln}^+)(\omega-\omega_{ln}^-),
\label{eqcomplexfreqsquared}
\end{equation}
where $\omega_{ln}^\pm=-\ii\gamma_{ln} \pm \sqrt{\omega_{ln}^2-\gamma_{ln}^2}$. Commonly, $\omega_{ln}^2 > \gamma_{ln}^2$ is valid. Near a resonance peak, we have $\omega \approx \omega_{ln} \approx \omega_{ln}^+$, the sum in \eqref{eq0power} is thus dominated by the term with $n'=n$, and we have
\begin{align}
         \Power_0(l,\omega)
				&\approx  \omega^2 M(\omega)  f(l,\omega)^2 (2l+1) \frac{1}{4\omega_{ln}^2}\frac{ R_{ln}^2(\robs) R_{ln}^2(r_s) }{( \omega-\omega_{ln})^2 + \gamma_{ln}^2}.\label{eq0powerapprox}
\end{align}
Equation~\eqref{eq0powerapprox} is very similar to equation~(54) in \citet{BirchKo2004}. We thus conclude that equation~\eqref{eq0power} can reproduce observational power spectra equally well as \citet{BirchKo2004}. An example zero-order power spectrum corresponding to one of the kernels computed in Section~\ref{secdeepkernels} is shown in Figure~\ref{figzeropowercrosscorr} (left panel).

\subsection{Zero-Order Cross-Covariance}
\label{seczeroordercrosscorr}
\begin{figure}

\includegraphics[keepaspectratio,width=0.5\columnwidth]{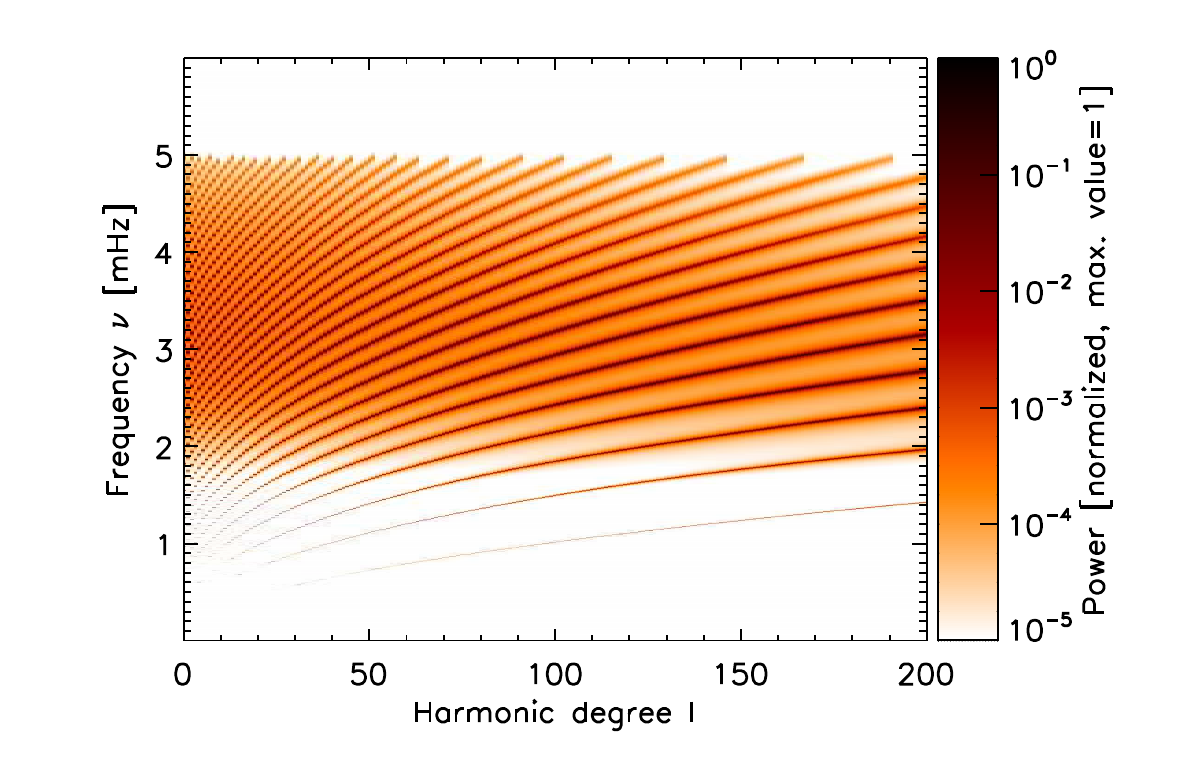}
\includegraphics[keepaspectratio,width=0.5\columnwidth]{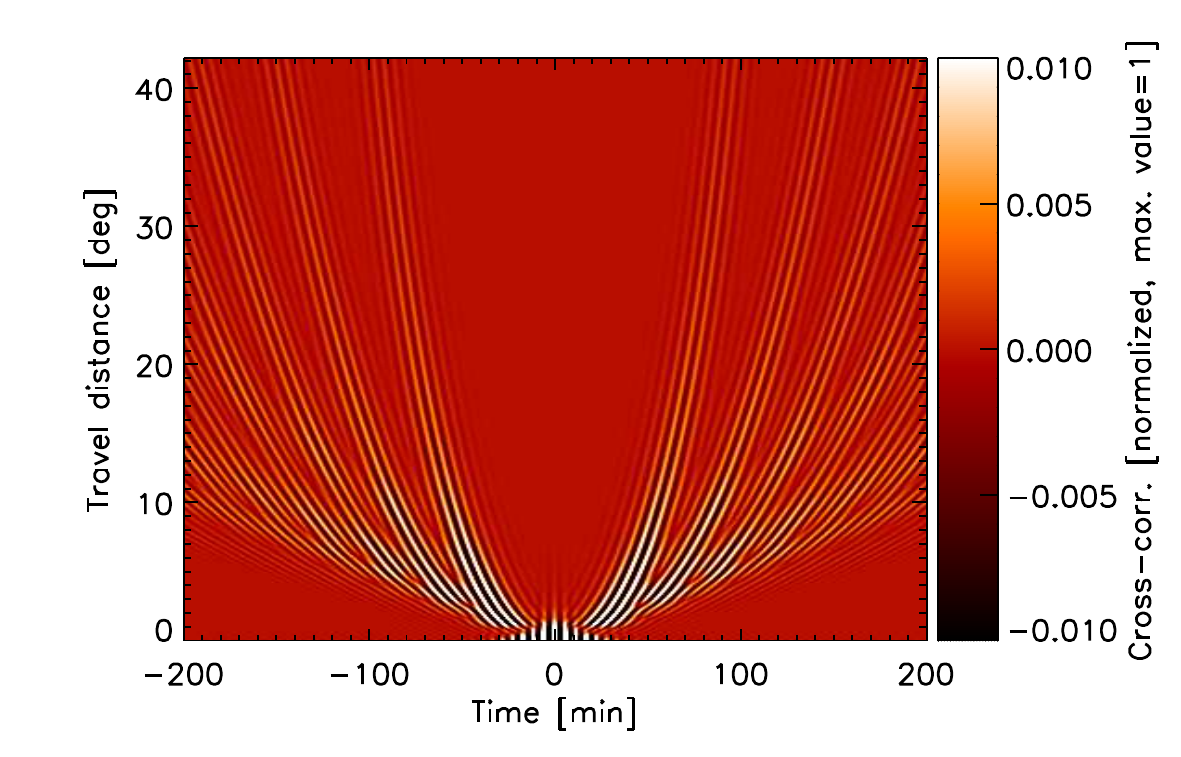}
\caption{Example zero-order power spectrum (left) and time-distance diagram (right). Both results were obtained with the same parameters as for kernel $K_1$, see Section~\ref{secdeepkernels}, but with $l\leq1000$ and with the same optical transfer function as in Section~\ref{secsanity} in order to better match observations. \label{figzeropowercrosscorr}}
\end{figure}

The following expression for the zero-order cross-covariance function $C_0$ is calculated in Appendix \ref{appendixcrosscorr}, similarly to the zero-order power spectrum, as an expectation value of the cross-covariance. It is also used as a reference cross-covariance, $\Cref=C_0$. We find
\begin{equation}
C_0(\br_1,\br_2,\omega) =  \frac{\omega^2}{4\pi}  M(\omega)  \sum_{lnn'} (2l+1) f(l,\omega)^2 R_{ln}(r_s) R_{ln'}(r_s)   \frac{ R_{ln}(r_1) R_{ln'}(r_2) P_l(\cos \Delta_{1,2}) }{(\sigma^{2*}_{ln}-\omega^2) (\sigma^2_{ln'}-\omega^2 )}, \label{eq0crosscorr}
\end{equation}
where $P_l$ is a Legendre polynomial of degree $l$, see Appendix \ref{appendixcrosscorr}. Given two identical radial locations of observation, $r_1=r_2=\robs$, the following relation between the cross-covariance and the power spectrum holds,
\begin{equation}
C_0(\br_1,\br_2,\omega) = \frac{1}{4\pi} \sum_l \Power_0(l,\omega) P_l(\cos \Delta_{1,2}) \label{eq0crosscorrpower}.
\end{equation}
Thus, if the zero-order power spectrum matches observations, this also applies to the zero-order cross-covariance function. An example zero-order time-distance diagram $C_0(\Delta_{1,2},t)$ corresponding to one of the kernels computed in Section~\ref{secdeepkernels} is shown in Figure~\ref{figzeropowercrosscorr} (right panel).

\section{SPHERICAL KERNEL FORMULA}

\label{seckernelformula}

\subsection{First-Order Problem}
\label{secfirstorder}

Weak flows result in a perturbation of the wave equation~\eqref{eqzeroorder} which we treat with first-order perturbation theory. All quantities $q$ are perturbed by an amount $\delta q$. Solving the first-order problem is necessary for obtaining an expression for the sensitivity functions, which represent a linear relationship between the travel-time shifts and the flow field as in equation~\eqref{eqkernelgoal}.

The equation for the single-scattering Born approximation to the perturbed wave equation in the presence of flows is (see \citealp{GB2002})
\begin{align}
(\cL + \delta \cL) [\bxi +\delta\bxi]  &= \bS + \delta \bS. \label{eqfirstorder}
\end{align}
\citet{Jackiewicz2007a} showed that the perturbation of the source and damping terms is negligible at first order, we thus assume $\delta \bS = \delta \Gamma = 0$. We also neglect the second order term $\delta \cL [\delta\bxi]$. At first order, solving equation~\eqref{eqfirstorder} is then equivalent to solving
\begin{align}
\cL[\delta\bxi] = -\delta \cL [\bxi].
\label{eqfirstordershort}
\end{align}
Following \cite{BG2007}, we assume that the perturbation to the wave operator is given by wave advection,
\begin{align}
\delta \cL [\bxi] = \delta \cW [\bxi] = -2\ii\omega \rho_0(r) \; \bv(\br) \bcdot \bnabla_\br  \left[\bxi(\br,\omega) \right],
\end{align}
where the index $\br$ in $\bnabla_\br[\cdot]$ denotes that $\bnabla$ works on the variable $\br$ of the function in square brackets.

As the structure of the first-order problem in equation~\eqref{eqfirstordershort} is identical to equation~\eqref{eqzeroorder}, we can use the Green's functions from equation~\eqref{eqgreens} to express its solution, replacing $\bxi$ by $\delta \bxi$ and $\bS$ by $-\delta \cL[\bxi]$ (see \citealp{GB2002}). The result is
\begin{align}
        \delta \xi_j(\brobs,\omega)  =  \int_\sun G^k_j(\brobs|\br,\omega)  \,  2\ii\omega \rho_0(r) \, \bv(\br) \bcdot \bnabla_{\br}  \left[\xi_k(\br, \omega) \right]  \, \id^3 \br  \label{eqdeltaxi}.
\end{align}
From the expression for the perturbation to the wave field in equation~\eqref{eqdeltaxi}, we can model the perturbation to the filtered Doppler signal via
\begin{align}
        \delta \Phi(\brobs,\omega)
				=  -\ii \omega \int_\sun \cG^k(\brobs|\br,\omega)  \,  2\ii\omega \rho_0(r) \, \bv(\br) \bcdot \bnabla_{\br}  \left[\xi_k(\br, \omega) \right]  \, \id^3 \br \label{eqdeltaPhi}.
\end{align}

\subsection{A General Kernel Formula}

\label{secgeneral}

In order to obtain a spherical formula for our sensitivity kernels, we first closely follow \cite{GB2002} and \cite{BG2007}. We need to first establish a linear relationship between the perturbation to the cross-covariance,
\begin{equation}
\delta C_0(\br_1,\br_2,\omega) = \EE \left[ \frac{2 \pi}{T} \, (\Phi+\delta\Phi)^*(\br_1,\omega) \,  (\Phi+\delta\Phi)(\br_2,\omega) \right] - \EE \left[ \frac{2 \pi}{T} \, \Phi^*(\br_1,\omega) \,  \Phi(\br_2,\omega) \right],
\label{eqdeltaC1}
\end{equation}
and the flow $\bv(\br)$, which can be written as
\begin{equation}
 \delta C(\br_1,\br_2,t) =\int_\sun  \bcurlyC(\br_1,\br_2,t;\br) \bcdot  \bv(\br) \, \id^3 \br \; .
\label{eqdeltaC}
\end{equation}
Equation~\eqref{eqdeltaC} then yields with equations \eqref{eqkernelgoal} and \eqref{eqtau}
\begin{equation}
\int_\sun  \, {{\bf K}(\br_1,\br_2; \br)} \bcdot \bv(\br) \, \id^3 \br
= \EE\left[\delta \taudiff(\br_1,\br_2)\right] 
= 2\pi \int_{-\infty}^\infty \Wdiff(\br_1,\br_2,\omega)  \, \delta C(\br_1,\br_2, \omega) \, \id t, \label{eqkernelderiv1}
\end{equation}
where the sensitivity kernel $\bK$ can be written as
\begin{align}
        {\bf K}(\br_1,\br_2; \br) 
                &= 2 \pi \int_{-\infty}^\infty \Wdiffstar(\br_1,\br_2,\omega) \,  \bcurlyC(\br_1,\br_2,\omega;\br)  \, \id \omega \label{eqkerneldef}
\end{align}
and where we have applied Parseval's theorem (see eq. \ref{eqparseval}, note also that $\Wdiff(\br_1,\br_2,t)$ is a real function).

In Appendix \ref{appendixgeneral}, we obtain an expression for $\bcurlyC$, see equation~\eqref{eqc1}, from which we deduce the general kernel formula
        \begin{align}
        {\bf K}&(\br_1,\br_2; \br)      = 4 \pi \, \rho_0(r)  \int_{-\infty}^\infty  M(\omega) \omega^3  \Wdiffstar(\br_1,\br_2,\omega)  \nonumber \\
        & \;\;\; \;\;\; \times  \bigg( \ii  \,  {\cG}^{k}(\br_2|\br,\omega)  \int_{S^2} \bnabla_{\br} [  G^{r}_k(\br|r_s, \Omega',\omega)] \, {\cG}^{r*}(\br_1|r_s, \Omega',\omega) \, \id  \Omega'   + (1 \leftrightarrow 2)^* \bigg) \, \id \omega, \label{eqkernelgeneral}
\end{align}
where the term abbreviated with $(1 \leftrightarrow 2)^*$ is identical to the previous term in the big round brackets except for complex conjugation and exchange of indices 1 and 2.

Equation~\eqref{eqkernelgeneral} is in fact very similar to equations (10) and (11) in \cite{BG2007}, which were found for Cartesian geometry. \cite{BG2007} used their equivalent of equation~\eqref{eqkernelgeneral} for an implemention into numerical code. As is pointed out in Section \ref{secnumerics}, it is computationally preferable in our spherical case to first perform the integral over $\omega$, before the spatial integral in equation~\eqref{eqkernelgeneral} is performed. This in turn requires us to expand the expressions involving the Green's functions $G^r_k$ and $\cG^k$ in \eqref{eqkernelgeneral}.

\subsection{The Specific Kernel Formula}
\label{secspecific}

Appendix \ref{appendixspecific} describes in detail how we obtain the following specific kernel formula from equation~\eqref{eqkernelgeneral},
\begin{align}
        \bK(\br_1,\br_2; \br)=& \sum_{j=(ln), i=(\bar l \bar n)} \; J_{ij}(\br_1,\br_2) \, \bZ^{ij}(\br_1,\br_2; \br)  + \Big( 1 \leftrightarrow 2 \Big)^* \label{eqcalckernel},
\end{align}
where
\begin{align}
        {\bZ^{ij}(\br_1,\br_2; \br)} &= \rho_0(r)   { \sum_{k=r,\theta,\phi} \cO_k^{\bar l \bar n} (\br)\Big[ P_{\bar l}(\cos \Delta_2)\Big]  {\bm \bnabla}_{\br} \Bigg[   \cO_k^{ln} (\br) \Big[  P_{l}(\cos \Delta_1)  \Big]     \Bigg]}  , \label{eqZij}   \\
{J_{ij}(\br_1,\br_2)} &= (2l+1)(2\bar l +1) R_{ln}(r_s) R_{\bar l \bar n}(r_2)\nonumber \\
			& \, \, \, \times { \sum_{n'}  R_{ln'}(r_s) R_{ln'}(r_1)   \int_{-\infty}^\infty  \frac{\ii \omega^3 \Wdiffstar(\br_1,\br_2,\omega)  M(\omega) \, f(l,\omega) f(\bar l,\omega)}{4\pi \, (\sigma^{2}_{ln}-\omega^2) (\sigma^{2*}_{ln'}-\omega^2)(\sigma^{2}_{\bar l \bar n}-\omega^2)} \, \id \omega  } .\label{eqHij}
\end{align}
We note that $\Wdiffstar$ is the only term which is not complex conjugated in the $(1 \leftrightarrow 2)^*$ term.

According to Equation~\eqref{eqcalckernel}, the kernel is given by a sum over all pairs of modes $(i,j)$. For each pair of modes $(i,j)$, the term $J_{ij}(\br_1,\br_2) Z^{ij}_d(\br_1,\br_2; \br)$ describes the advection-induced scattering of mode $j$ by a flow in direction $d \in \{ r, \theta, \phi\}$, which results in a perturbation of the travel time through a coupling with mode $i$. The quantity $J_{ij}$ describes the observational signature of the coupling between mode $i$ and $j$ which depends on the background model and the travel distance only. The term $Z_d^{i,j}$ describes the observable strength of the scattering and the coupling, which depends on the density-weighted scattering location in the solar interior relative to the observation points.

\section{NUMERICAL IMPLEMENTATION}
\label{secnumerics}

Equation~\eqref{eqcalckernel} is used for the numerical implementation of the kernel formula. In order to reduce computation time, the coupling between $n$ and $n'$ was neglected in the evaluation of the kernel formula as well as in the computation of the zero-order power spectrum and the zero-order cross-correlation. Thus, only terms with $n'=n$ were taken into account in equations \eqref{eq0power}, \eqref{eq0crosscorr}, and \eqref{eqHij}. For a particular harmonic degree $l$, the coupling between two distinct ridges in the power spectrum, $n'\neq n$, is expected to be small because the frequencies of corresponding modes differ by an amount much greater than the linewidth. However, this should be checked for any specific application. If necessary, the full formulas can be used.

Equation~\eqref{eqcalckernel} shows the complexity of the computation. If we denote the number of grid points in $r, \theta, \phi$ with $N_r,N_\theta, N_\phi$ and the number of modes used in the model calculation with $N_i$, then the numerical performance of the evaluation of the sum over $(i,j)$ in equation~\eqref{eqcalckernel} scales with 
\begin{equation}
\text{cost}_K = 	N_r N_\theta N_\phi \; N_i^2 .
\label{eqcostK}
\end{equation}
The computation of $J_{ij}$ as defined in equation~\eqref{eqHij}, which can be performed prior to taking the sum over $(i,j)$, scales with
\begin{align}
\text{cost}_{J_{ij}} &= N_i^2 N_\omega N_{n'} \label{eqcostHij}
\end{align}
where $N_\omega$ is the number of points used in the $\omega$ grid and $N_{n'}$ is the number of ridges used in the computation of the line asymmetry in $ J_{ij}$. The cost of computing $\bZ^{ij}$ is included in the evaluation of the sum over $(i,j)$ in equation~\eqref{eqcalckernel}.

Alternatively, it is possible to follow \cite{BG2007} by using equation~\eqref{eqkernelgeneral} for the implementation (see also \citealp{Burston2015}). In this case, one evaluates equation~\eqref{eqconvolutionGG} for every frequency and then performs the frequency integral in equation~\eqref{eqkernelgeneral}. The numerical cost would then scale with
\begin{equation}
\text{cost}_{\text{alternative}} = N_r N_\theta N_\phi \; N_\omega \; N_i.
\label{eqcostalternative}
\end{equation}

From equations \eqref{eqcostK} and \eqref{eqcostalternative} ($\text{cost}_{ J_{ij}} \leq  \text{cost}_K$ for reasonable examples), it is clear that the alternative procedure is faster if $N_\omega < N_i$, which is the case when high-degree modes are considered as in \citet{BG2007}. If we consider low degree modes, e.g. for calculating kernels for large travel distances, this is no longer appropriate because low degree modes have considerably smaller damping rates. In order to correctly evaluate the integral over $\omega$ in equations \eqref{eqkernelgeneral} or \eqref{eqcalckernel}, we therefore have to use a very fine frequency resolution, see also Section~\ref{secmeridionalflowkernels}. We thus have $N_\omega >> N_i$ and the approach taken in this paper is computationally preferable.

\section{SANITY CHECK: COMPARISON WITH CARTESIAN GEOMETRY}

\label{secsanity}

In this section, we perform a first sanity check of the derived kernel formula and its implementation. We perform two kernel computations for an example travel distance of $\Delta=10\,\mathrm{Mm}$ using the same set of model parameters, one with our spherical code and one with the existing Cartesian code developed by \citet{BG2007}, and we compare the results.

\subsection{Sanity Check: Model Parameters}
\label{secsanityparameters}

We consider the $f$-mode example in \citet{BG2007}, which was also studied by \citet{Jackiewicz2007a}. In order to facilitate the comparison, a number of changes were made to the model parameters. 
As our spherical code does not take the line asymmetry in the power spectrum into account (see Section \ref{secnumerics}), both Cartesian and spherical computations were performed using only the $f$-mode in the normal mode summation. \citet{BG2007} also used modes with higher radial orders $n>0$ in their computation, where the cross-talk between, e.g., the $p_1$ and the $f$-mode has a contribution to the power spectrum. 
As the spherical code uses a grid in $l$, the resolution in $k$ in the Cartesian code was changed to correspond to our requirement of $\Delta l =1$. In the spherical code, we use $l_\text{min}=139$ and $l_\text{max}=2053$, the corresponding choice is made for $k$ in the Cartesian code. Finally, we employ a frequency filter such that only frequencies $\nu$ between 2 mHz and 4 mHz are taken into account in the computations. 
For the sake of comparison with the kernels presented in the current paper, we included an additional multiplicative factor of $\omega^2$ to the source covariance in the Cartesian code of \citet{BG2007}.  This factor of $\omega^2$ comes from the difference in the source covariance models employed. 
The source type was further changed (we here assume vertical momentum sources, which corresponds to omitting the radial derivatives in equation~(51) in \citealp{BirchKo2004}) and we use a different model for the optical transfer function (\citealp{BG2007} actually used $\alpha=1\,\rm{Mm}$ instead of the mentioned $\alpha=1.75\,\rm{Mm}$). We note also that the code of \citet{BG2007} uses eigenfunctions which were computed in Cartesian geometry, while the eigenfunctions used in the spherical code were computed in spherical geometry.

\subsection{Sanity Check: Comparison of Kernel Results}

\begin{figure}

\includegraphics[keepaspectratio,width=0.5\columnwidth]{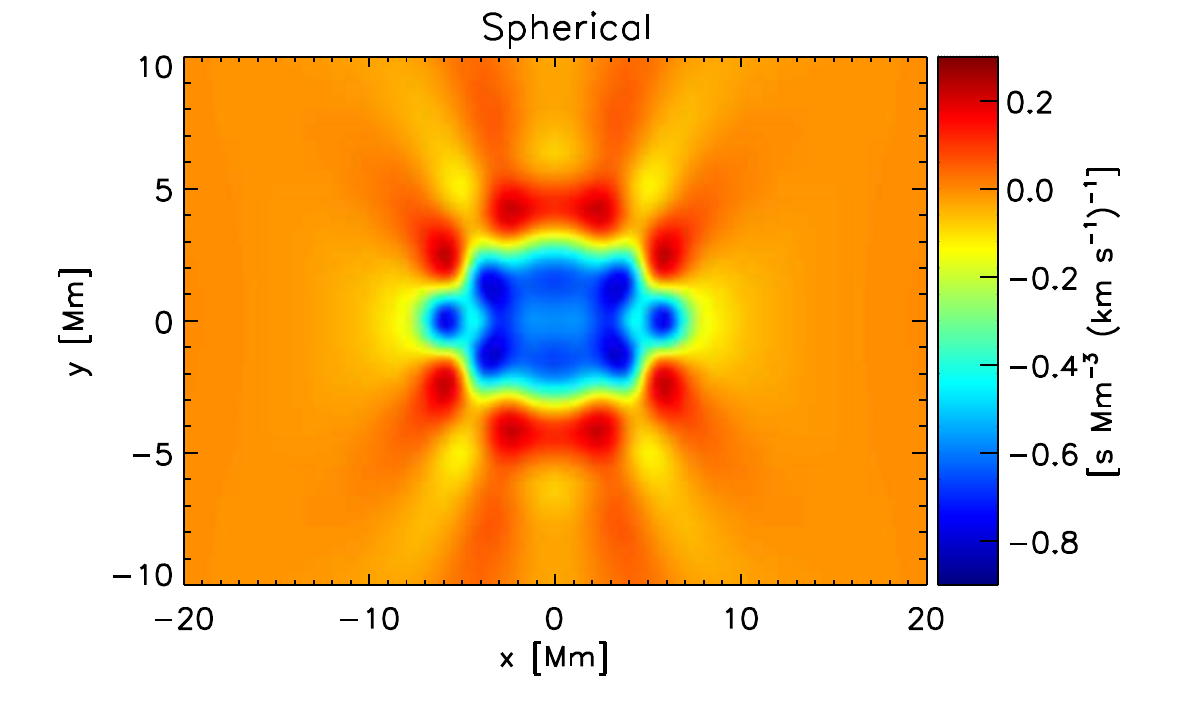} 
\includegraphics[keepaspectratio,width=0.5\columnwidth]{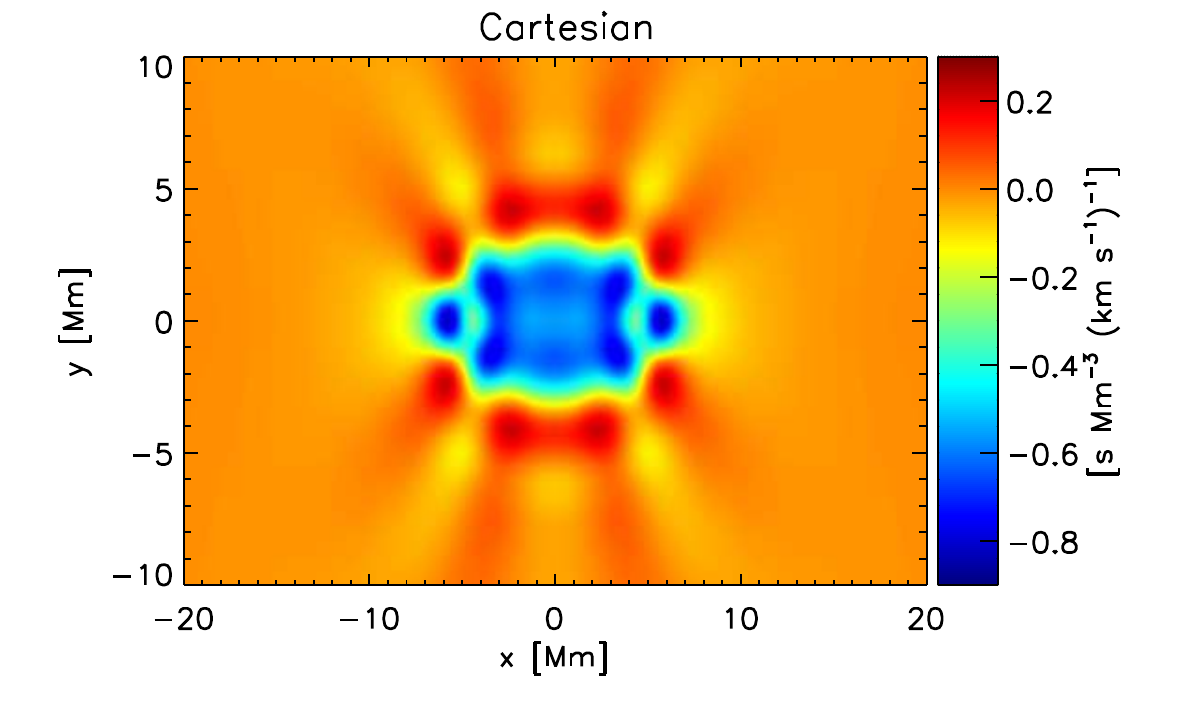}

\caption{Sensitivity of travel-time differences to zonal flows. Displayed are horizontal cuts at the source depth through $K_\phi$ (left, spherical code) and $K_x$ (right, from the Cartesian code used by \citealp{BG2007}). The two observation points are located on the equator ($y=0$) at $x=\pm5\;\rm{Mm}$. \label{figfmodehorizontal} }
\end{figure}

Figure~\ref{figfmodehorizontal} shows horizontal cuts through the kernels for zonal flows,  $K_\phi$ (from our spherical code) and $K_x$ (from the Cartesian code used by \citealp{BG2007}) at the source depth. 
The results are in good agreement. Regions of positive and negative sensitivity are found at the same locations and the magnitude of the kernels is of similar order, which we evaluate quantitatively in the following.

\begin{figure}
\includegraphics[keepaspectratio,width=0.5\columnwidth]{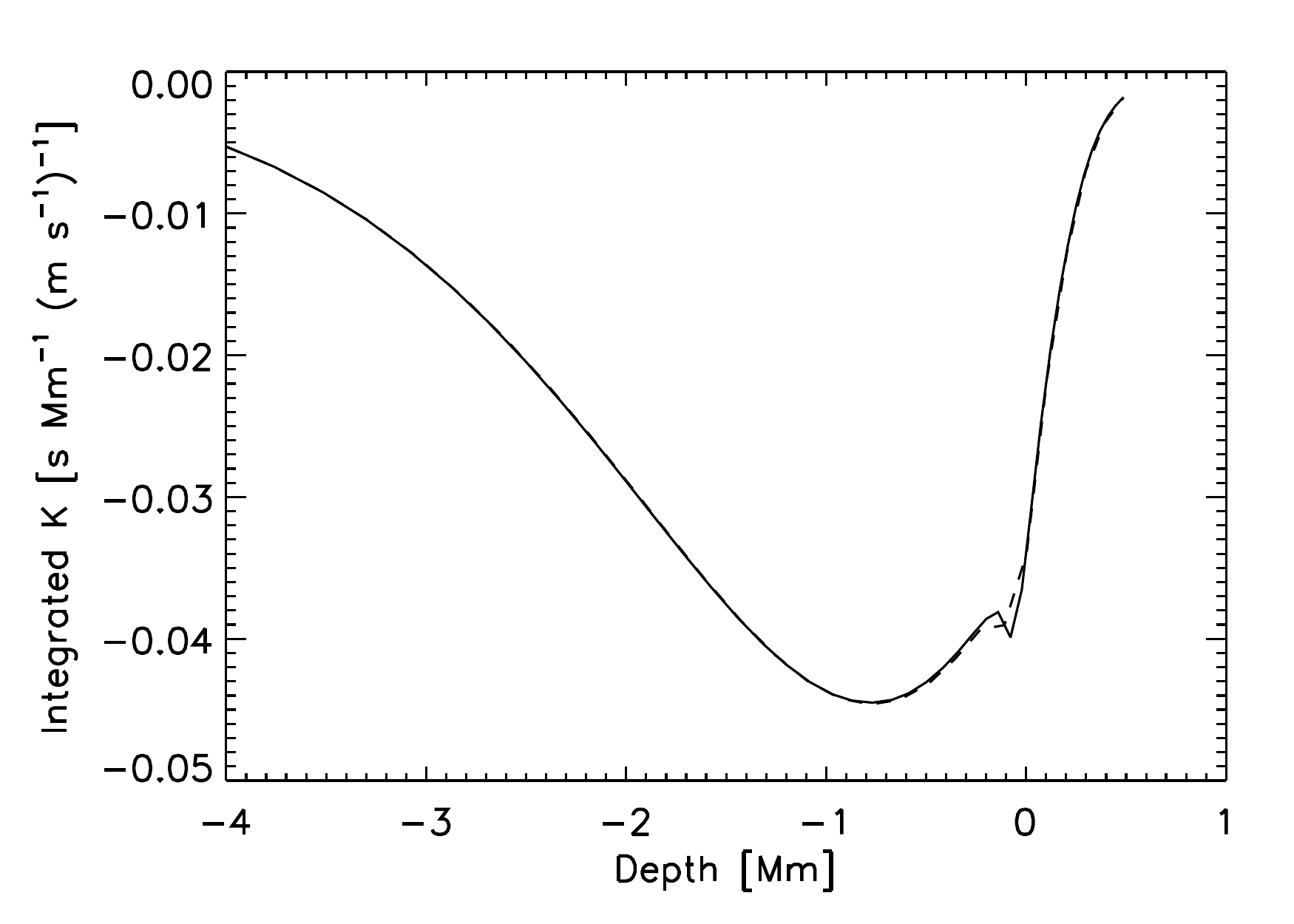}
\includegraphics[keepaspectratio,width=0.5\columnwidth]{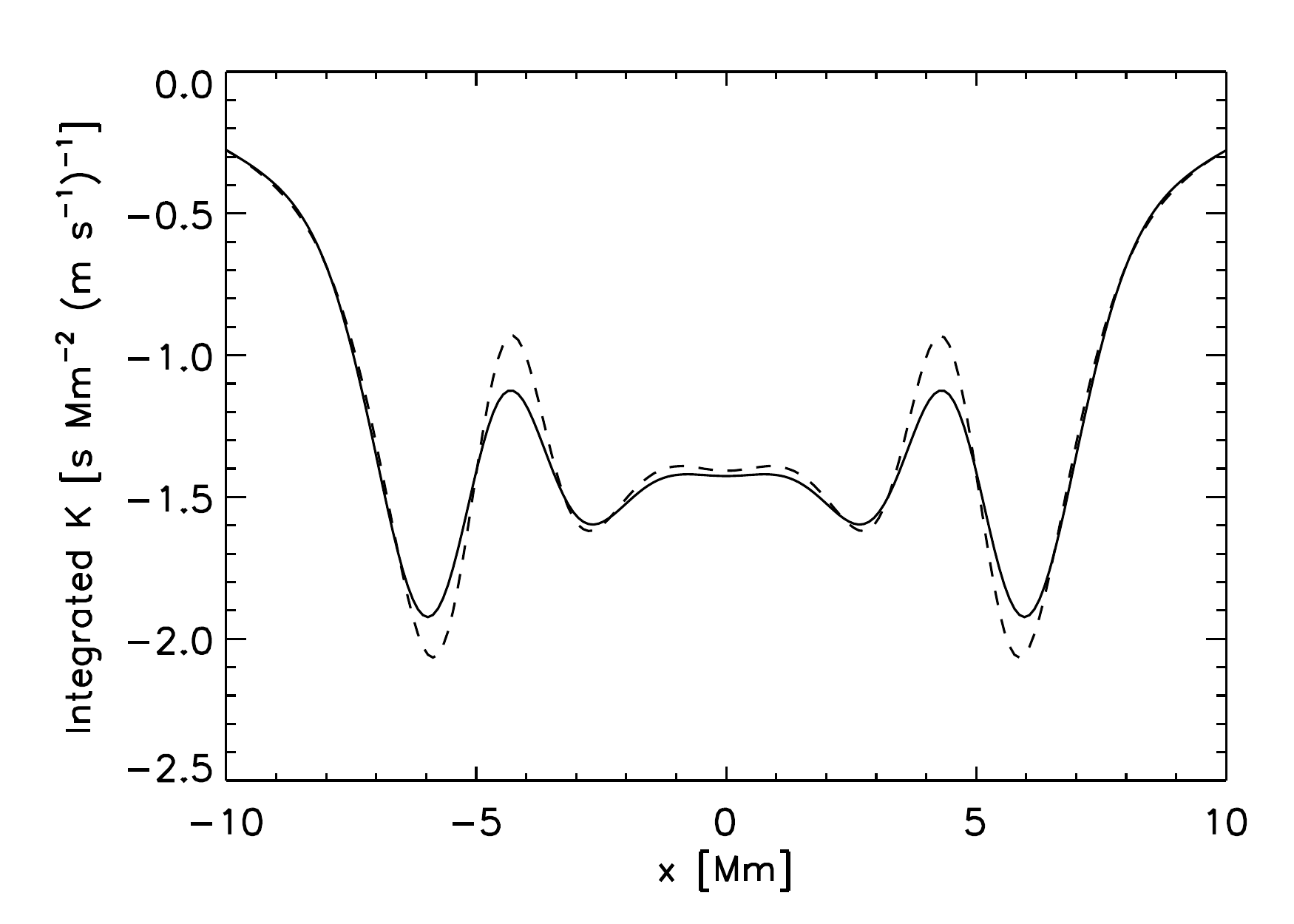}
\caption{Comparison of integrated travel-time sensitivity to zonal flows from the spherical code ($K_\phi$, solid lines) and from the Cartesian code ($K_x$ in \citealp{BG2007}, dashed lines). The left panel shows the horizontally integrated sensitivity kernels as functions of depth, the right panel shows cuts at the equator through the radial integrals of the same kernels as a function of $x$ or $\phi$, respectively.\label{figfmodeintegrated}}
\end{figure}

In Figure~\ref{figfmodeintegrated} and Table~\ref{tablesanity}, the magnitudes of the two kernels are compared in more detail. The left panel of Figure~\ref{figfmodeintegrated} shows the horizontally integrated sensitivity to zonal flows as a function of depth from the spherical code ($K_\phi$, solid line) and from the Cartesian code ($K_x$, dashed line). The right panel shows cuts at the equator through the radial integrals of the same kernels as a function of horizontal distance $x$ in the W-E direction. The horizontal integrals are in very good agreement, the relative difference of the negative peak values is found to be about 0.28~\%. The total integrals of the kernels, which correspond to the sensitivity to a uniform flow field, differ by only about 0.14~\%.

We thus conclude for the given example that the spherical and the Cartesian approach give very similar travel-time sensitivities to large-scale flows and that the average magnitude of the spherical kernel is thus plausible.

In the radial integrals presented in the right panel of Figure~\ref{figfmodeintegrated}, the Cartesian kernel shows an oscillatory behaviour with higher amplitude compared to the spherical one, for which the maximum absolute value in the radial integral in Figure~\ref{figfmodeintegrated} (right panel) is lower by about 6.9~\% at the peaks. Similarly, the maximum absolute value of the spherical kernel is lower by 4.9~\% compared to the Cartesian one.

As a reason for the remaining differences in the two kernel results, we exclude slight differences in the power spectra due to differences in the damping model with the following test: 
Correcting the power spectrum from the spherical code for the ratio of the two $\omega$-integrated power spectra yields a slight change in the mean sensitivity but the more oscillatory behaviour of the Cartesian kernels is not reproduced. Further possible reasons include the approximations made in Cartesian geometry and the differences in the eigenfunction computation.

\begin{deluxetable}{ccccccc}
\tabletypesize{\scriptsize}

\tablecolumns{7}
\tablecaption{Key Characteristics of Kernels for the Sanity Check in Section \ref{secsanity} \label{tablesanity}}
\tablewidth{0pt}
\tablehead{
\colhead{Kernel} & \colhead{$\max(K_\phi)$} & \colhead{$\min(K_\phi)$} & \colhead{$\int K_\phi \;\id^3\br$} & \colhead{$\min(r^2\int K_\phi \; \id \Omega)$} & \colhead{Mean $\nu$} & \colhead{$\nu$ at max. power} 
\\
& \colhead{[$\rm{s}^2\;\rm{Mm}^{-3}\,\rm{km}^{-1}$]} & \colhead{[$\rm{s}^2\;\rm{Mm}^{-3}\,\rm{km}^{-1}$]} & \colhead{[$\rm{s}^{2}\, \rm{m}^{-1}$]} & \colhead{[$\rm{s}^2\;\rm{m}^{-2}$]} & \colhead{[mHz]} & \colhead{[mHz]}
} 
\startdata
\citet[Fig.~2]{BG2007} & 1.431 & -4.421 & -0.1887 & -0.1067 & 2.77 & 2.70 \\
Cartesian $f$-mode example & 0.236 & -0.848 & -0.1193 & -0.0446 & 2.55 & 2.41 \\
Spherical $f$-mode example & 0.233 & -0.806 & -0.1191 & -0.0445 & 2.53 & 2.27 
\enddata
\tablecomments{For the $f$-mode kernel presented in \citet[Fig.~2]{BG2007}, as well as for the Cartesian and spherical kernels presented in this paper, we show the maximum (second column) and minimum (third column) values, as well as the total integral (fourth column), the minimum value of the horizontal integral (fifth column), the power-weighted mean frequency (sixth column), and the frequency at maximum wavenumber-integrated power (seventh column). The values are for $K_x$ or $K_\phi$, respectively.}
\end{deluxetable}

\subsection{Sensitivity of Kernels to the Power Spectrum}

Born approximation sensitivity kernels respond sensitively to changes in the power spectrum (e.g., \citealp{BirchKo2004}). For interpreting data, model power spectra must be well matched to those from data. 
In order to demonstrate how kernels respond to such changes, we note that the extremal value of the Cartesian $f$-mode kernel computed in this paper differs by a relatively large factor of about five from that presented in \citet[Fig.~2a]{BG2007}. This factor is exclusively due to the different choice of parameters for the kernel computation in this paper (see Section \ref{secsanityparameters}) compared to the parameters used by \citet{BG2007}, especially the change of source type and a different optical transfer function, and the resulting changes in the power spectrum. See also Table~\ref{tablesanity} for some key values for the two Cartesian kernels compared here.

\section{EXAMPLE KERNELS FOR LARGE TRAVEL DISTANCES}

\label{secdeepkernels}

The main application for the spherical kernels presented in this paper is to measure large-scale flows in the deep solar interior. In this section, we test example kernels of such an application for self-consistency and we evaluate the dependence of the kernels on the filters applied in the data analysis procedure. As an example, we consider a travel distance of 42 degrees. In the ray approximation, this corresponds to a lower turning point near the bottom of the convection zone (e.g., \citealp{Zhao2013}, \citealp{Rajaguru2015}).

\subsection{Testing Kernels in the Linear Regime}
\label{sectesting}

In this section, we test our results for self-consistency and follow the idea of \citet{RGB2006} by considering a model Sun rotating uniformly with a small angular frequency $\Omega$. In this case, we can analytically obtain a perturbed cross-covariance and thus reference values for travel times. Following \citet{Woodard1989}, we neglect the effect of the Coriolis force on eigenfunctions. Furthermore, the first-order effect of rotation on eigenfunctions due to advection vanishes according to \citet{Woodard1989} when the rotation is uniform. We thus purely consider the effect of rotation on mode frequencies, which are shifted in the non-rotating frame of the observer by (see, e.g., Section~3.8 and eq.~[3.360] in \citealp{Aerts2010})
\begin{equation}
	\delta\omega_{lmn} = m \beta_{nl} \Omega\label{eqmodeshift}
\end{equation}
due to advection and Coriolis forces. 
This frequency shift can be introduced in the cross-correlation if equation~\eqref{eq0crosscorr} is rewritten in a more general form,
\begin{equation}
C_0(\br_1,\br_2,\omega) = \sum_{lmn} f(l,\omega)^2 d_{lmn}(\br_1,\br_2,\omega), \label{eqcrosscorr0rewritten}
\end{equation}
where we have taken into account only terms with $n'=n$ (see Section 5). Equation~\eqref{eqcrosscorr0rewritten} becomes in the case of uniform rotation using Equation~\eqref{eqmodeshift},
\begin{equation}
\COmega(\br_1,\br_2,\omega) = \sum_{lmn} f(l,\omega)^2 d_{lmn}(\br_1,\br_2,\omega-\delta\omega_{lmn}), \label{eqcomega}
\end{equation}
where the filter was applied in the observer's frame.

\begin{figure}
\centering{
\includegraphics[width=0.7\columnwidth]{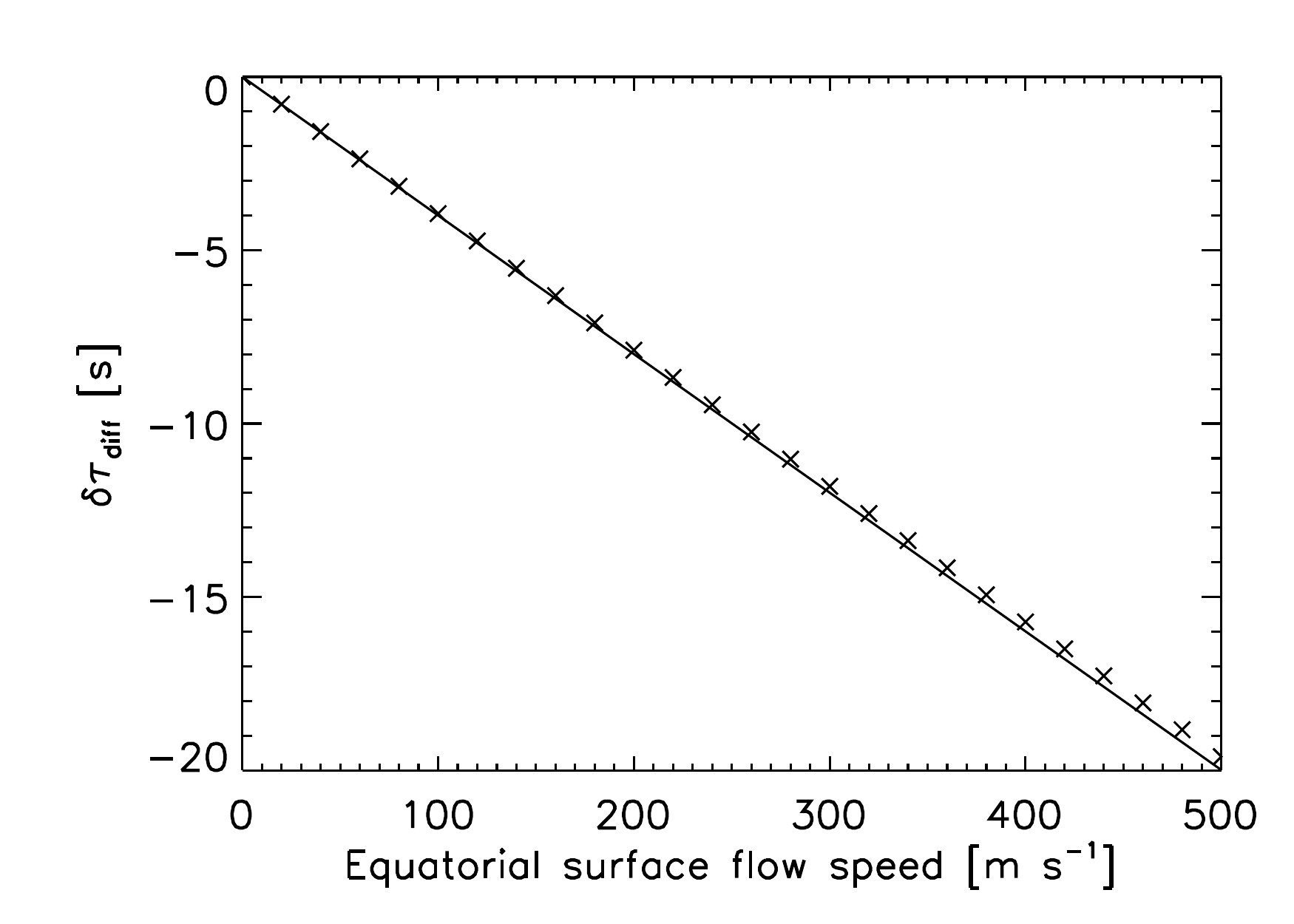}%
}
\caption{Perturbations to the travel-time difference due to uniform rotation as a function of the corresponding equatorial surface flow speed. The observation points are assumed to be aligned in the W-E direction on the equator at a distance of $42\degr$. We show forward-modelled travel-time differences from the kernel (solid line) and from an analytically obtained perturbed cross-covariance function (crosses).
\label{figcomegatest}}
\end{figure}

Figure~\ref{figcomegatest} depicts travel times obtained analytically using equation~\eqref{eqcomega} as a function of the equatorial surface flow speed $v=\Omega\,R_\sun$. For the computation of each $\COmega$, the observation points are assumed to be aligned in the W-E direction on the equator. Travel time differences $\tauomega$ in a westward minus eastward sense (W-E, crosses in Figure~\ref{figcomegatest}) are then fitted to each analytically-obtained cross-covariance function using equation~\eqref{eqtau}. A rotation with angular frequency $\Omega$ corresponds to a flow field of $\bv(\br)=\Omega\,r\sin\theta\,\unit^{(\phi)}(\br)$, which is used for obtaining the Born approximated travel times for such a flow (solid line, $\taukernel$) from equation~\eqref{eqkernelderiv1}. The travel times $\taukernel$ were obtained from a kernel which was computed using the same parameters as for $\comega$. Apart from the W-E-orientation, the parameters used for producing Figure~\ref{figcomegatest} are identical to those used for kernel $K_{4}$ (see Section~\ref{secmeridionalflowkernels} and Table~\ref{tablebigkernels}).

The travel times predicted from the kernel show a linear behaviour with increasing flow speed, while the analytically obtained travel times show a non-linear behaviour (see also \citealp{Jackiewicz2007a}). For small rotation rates, which correspond to small-amplitude uniform flows, linearity is a good approximation.

If the kernel is a good linear approximation to the effect of the flow field on the travel times, both $\tauomega$ and $\taukernel$ will have the same derivative with respect to $v$ at $v=0$. In order to obtain a quantitative measure for the accuracy of a kernel computation, we thus evaluate
\begin{equation}
	a = \frac{\left( \partial (\taukernel)/\partial v\right)_{v=0} }{ \left(\partial (\tauomega)/\partial v\right)_{v=0} }-1. \label{eqa}
\end{equation}
This value is a quantitative test of the numerical resolution and the extent of the spatial and frequency grids. 
For the example shown in Figure~\ref{figcomegatest}, we have $a=0.012$. Our kernel therefore gives an average sensitivity in the linear regime which is 1.2~\% larger than would be expected from the analytically obtained perturbed cross-covariance. 
At $v=500\,\rm{m}\,\rm{s}^{-1}$, the travel time obtained from the kernel is 1.8~\% larger than the one from the analytically obtained cross-covariance function. The assumption of small flows, equivalent to a linear relationship between $\delta\tau$ and $\bv$, is thus valid to a sufficient degree for flow speeds up to $500\,\rm{m}\,\rm{s}^{-1}$ (see Fig.~\ref{figcomegatest}) for this example.

\subsection{Meridional Flow Kernels for Different Filters and $\Delta=42\degr$: Model Parameters}
\label{secmeridionalflowkernels}

\begin{figure}%
\centering{

\includegraphics[keepaspectratio,width=0.3\textwidth]{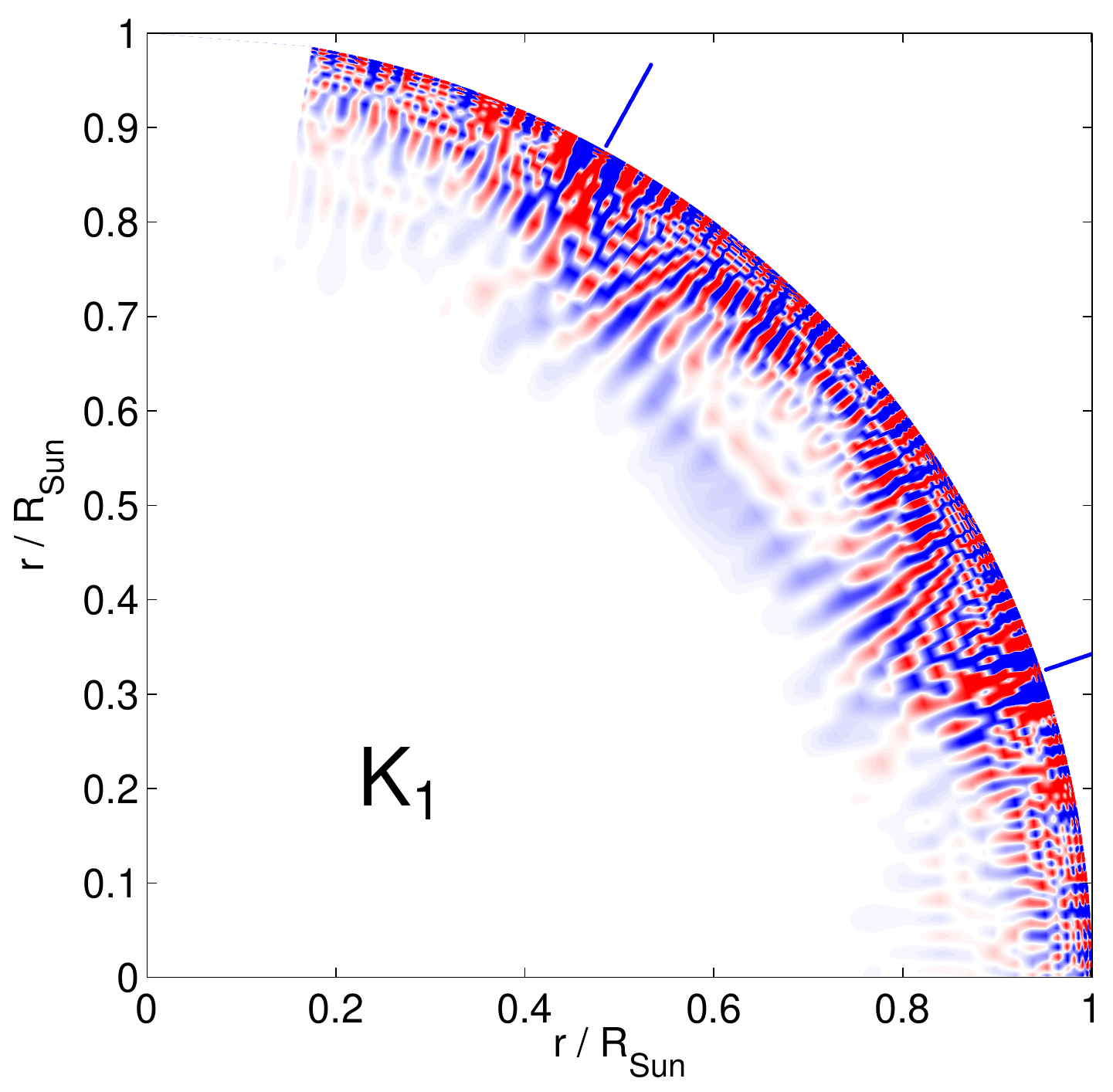}
\includegraphics[keepaspectratio,width=0.3\textwidth]{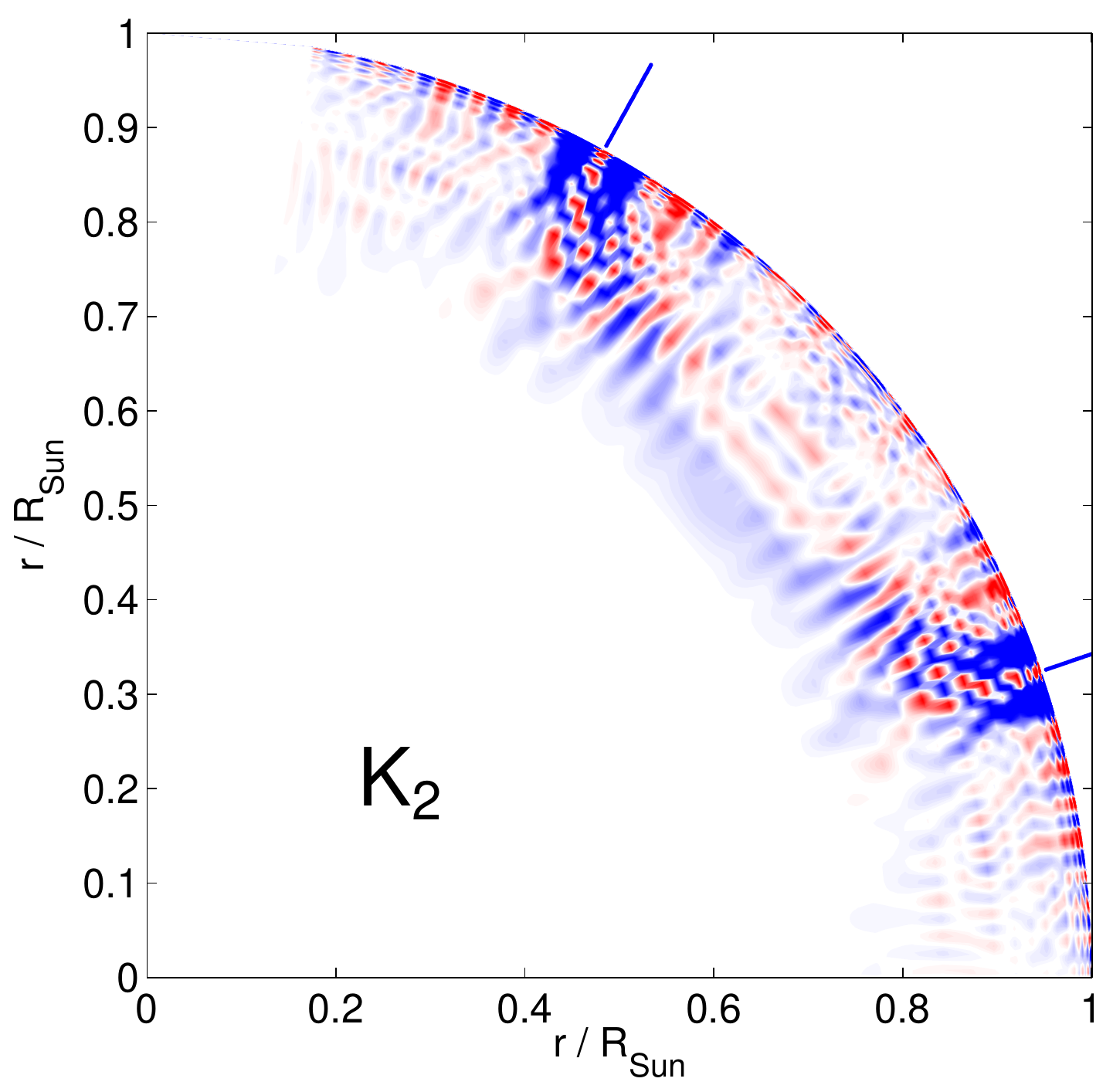}
\includegraphics[keepaspectratio,width=0.3\textwidth]{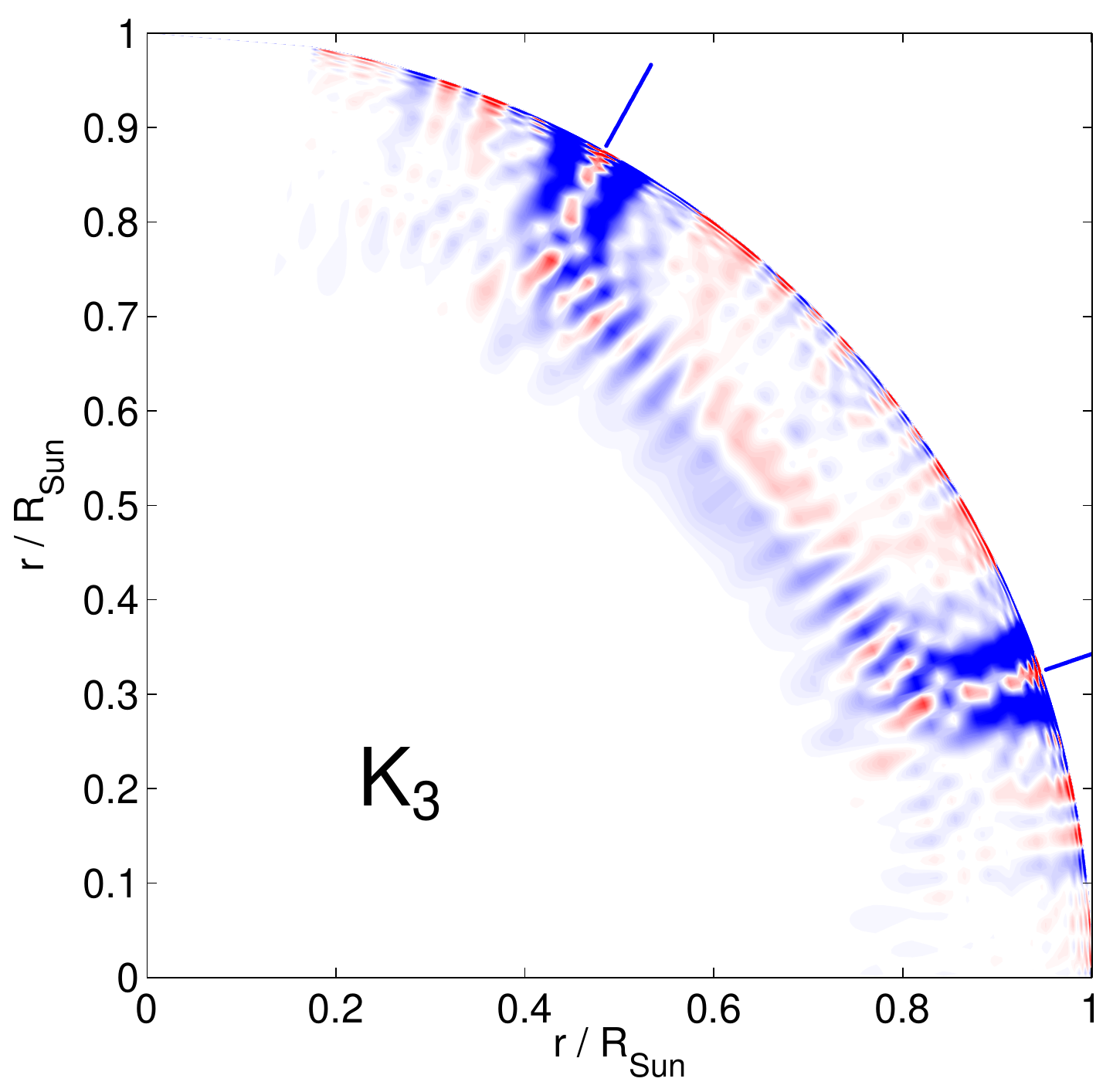}

\includegraphics[keepaspectratio,width=0.3\textwidth]{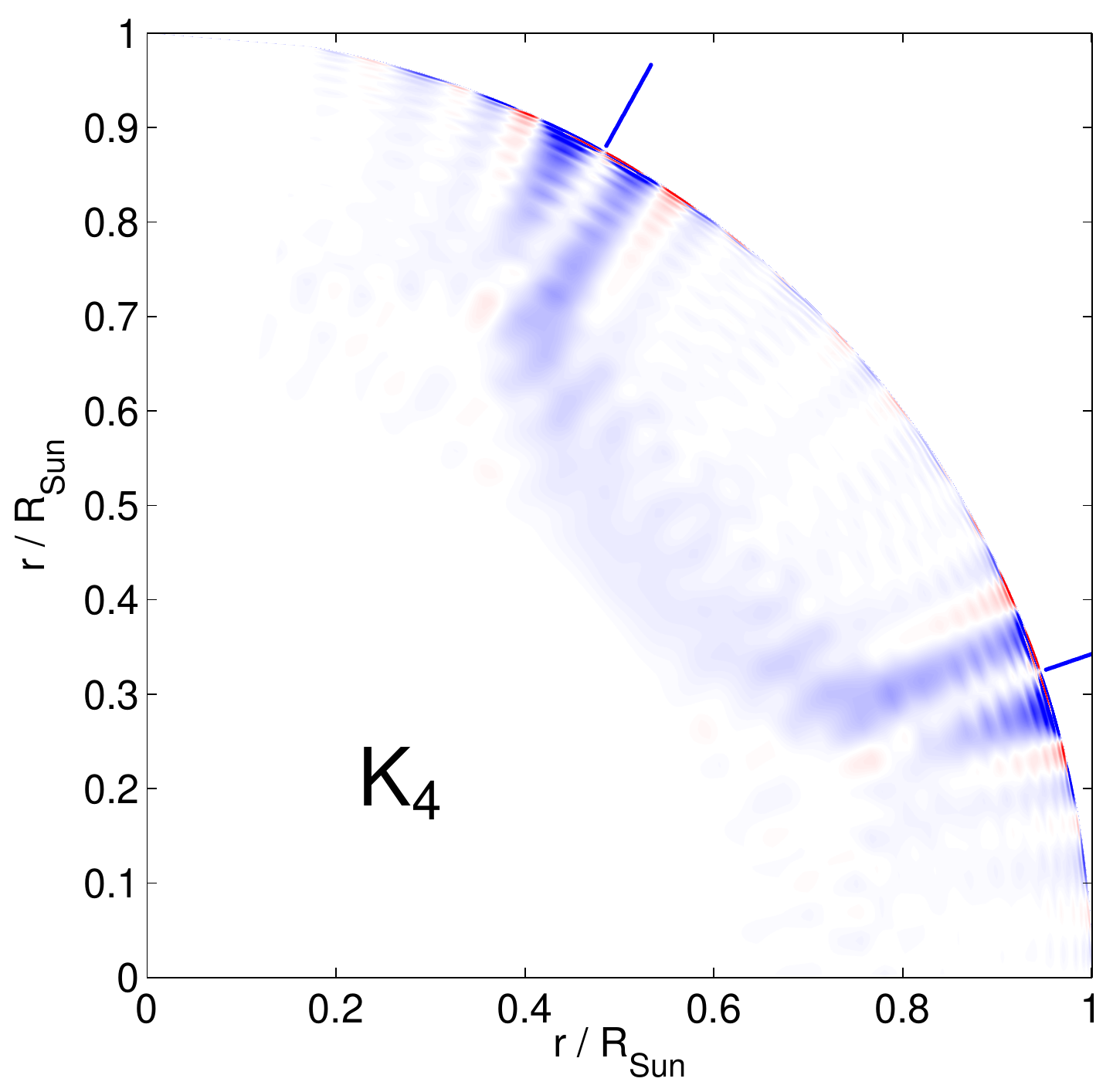}
\includegraphics[keepaspectratio,width=0.3\textwidth]{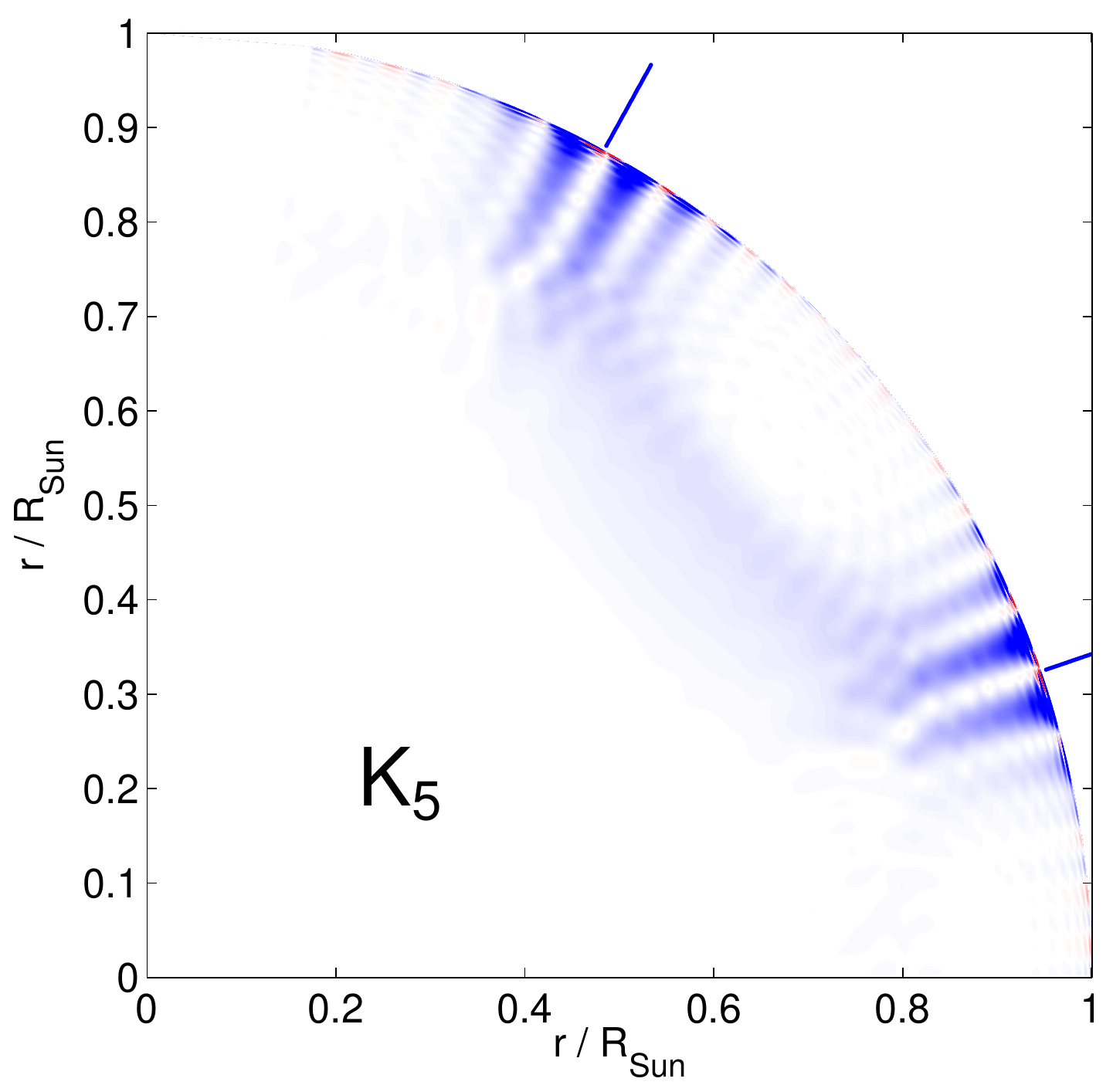}
\includegraphics[keepaspectratio,width=0.3\textwidth]{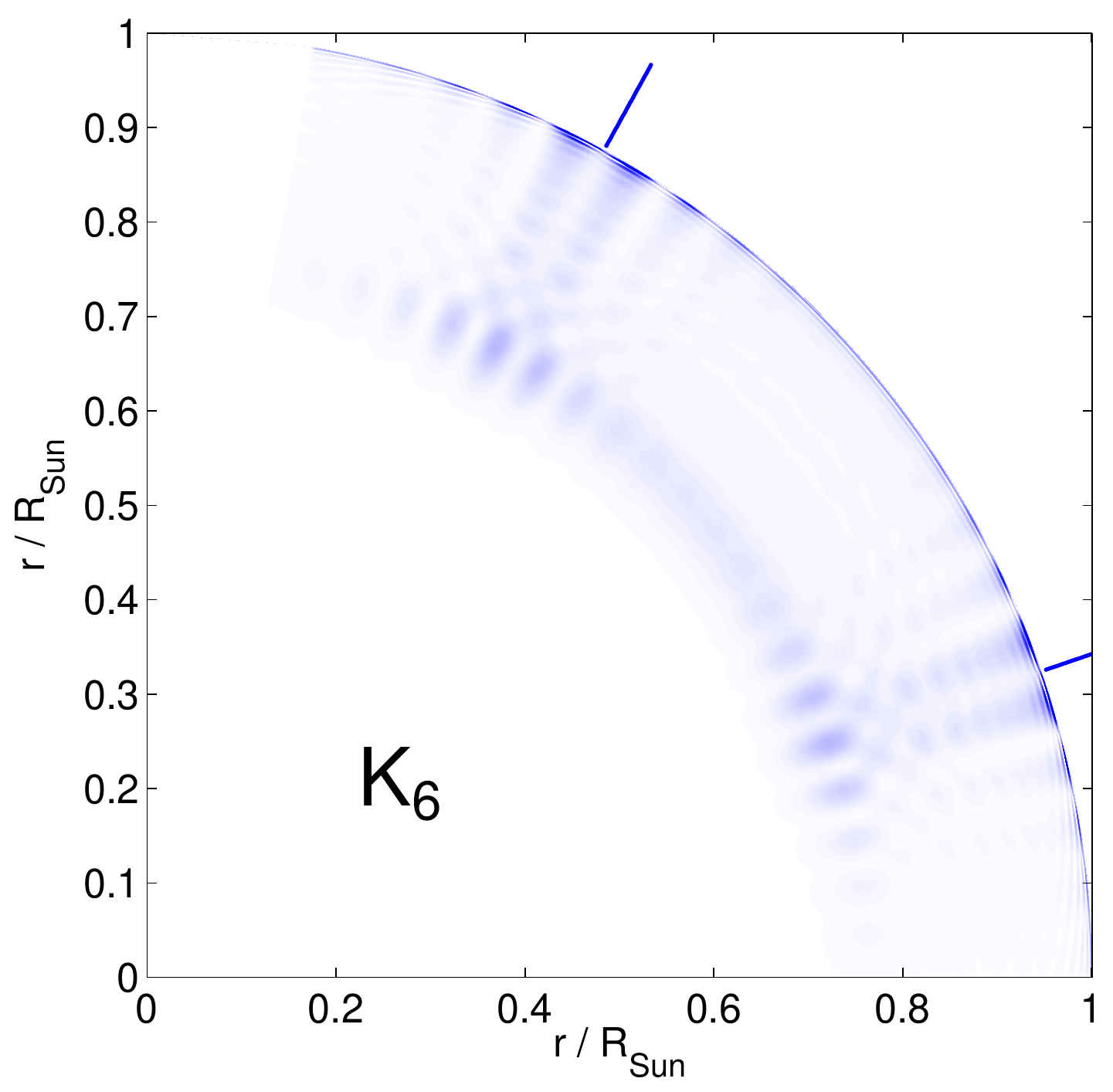}

\includegraphics[keepaspectratio,width=0.35\textwidth]{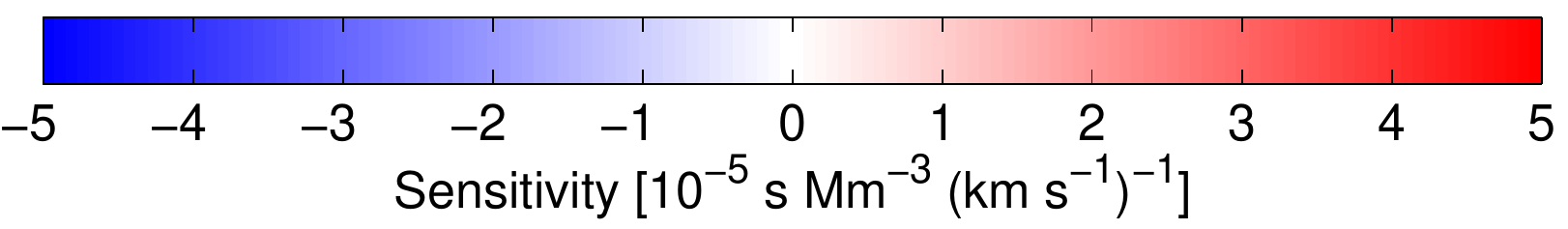}

}

\caption{
Travel-time sensitivity kernels for meridional flow, $K_\theta$, for a travel distance of $\Delta=42\degr$. Displayed are vertical cuts at the central meridian through kernels computed with six different filters, which are summarized in Table~\ref{tablebigkernels} and described in the text. The locations of the observation points are marked with blue bars.
\label{figbigkernels}}

\end{figure}

We calculate kernels for point-to-point southward minus northward (S-N) travel time differences with a travel distance of $\Delta=42\degr$ and observation points which are centered at a latitude of $40\degr$ and located along the central meridian ($\phi=0\degr$).
Figure~\ref{figbigkernels} displays vertical cuts at the central meridian through the example kernels. 
First, we show kernels for which the mode summation was truncated at $l\leq 170$, $l\leq99$, $l\leq 79$ (top row from left to right), and $l\leq 49$ (bottom left), respectively. For these kernels, apart from the truncation, no additional filters were applied. In addition, we show one kernel which was computed with a phase-speed filter (bottom right, filter no. 9 in \citealp{Kholikov2014}) and one kernel for which the modes were filtered with a Gaussian at a central angular degree of $l_0=45$ (bottom center), which corresponds roughly to the central mode present in the phase-speed filter (see also \citealp{Kholikov2014}). 
Table~\ref{tablebigkernels} summarizes some key characteristics of the kernels shown in this section.\notetoeditor{If possible, please place Table~\ref{tablebigkernels} near Figure~\ref{figbigkernels} for better reference.}

The following parameters are the same for all kernels. As in \citet{Hartlep2013}, we assume a source depth of $r_s=R_\sun -150\,\text{km}$ and an observation height of $r_1=r_2=\robs=R_\sun +300\,\text{km}$.  
The window function $h(t)$ used in the travel-time fit in equation~\eqref{eqW} was chosen to select the first bounce, i.e. it is equal to 1 for $70\,\text{min} \leq t \leq 120\,\text{min}$ and 0 otherwise for the given travel distance. For the examples shown here, we do not use an optical transfer function. Damping rates were supplied by Jesper Schou (2006, private communication).

While the selection of the above parameters is rather straightforward, the choice of the spatial and frequency grids turns out to be more complicated. In order to guarantee a good accuracy, e.g., $|a|\leq 0.05$, the grids in the spatial and frequency domains have to be sufficiently fine. The resolution of the spatial grid has to be at least as fine as the critical sampling of 
$\Delta \theta_\text{crit}=(1/2) \, \pi / (2 l_\text{max})$ 
in the horizontal direction, where $2l_\text{max}$ is a rough estimate of the angular degree of the finest horizontal structure in the kernel originating from the multiplication of two Legendre polynomials with maximum degree $l_\mathrm{max}$. We calculate the example kernels presented in this section in a volume which covers colatitudes from 10 to 90 degrees, longitudes from $-30$ to $+30$ degrees, and depths from $0.7R_\sun$ (in some cases $0.6R_\sun$) to the top of the model at around $R_\sun+500\,\text{km}$. We use 301 (or, if necessary to obtain an accuracy of $|a|<0.05$, 601) grid points in colatitude, 227 grid points in longitude, and 75 grid points in depth. This corresponds to a horizontal resolution which is roughly twice as fine as the critical sampling. 

The frequency grid is constrained by the low damping rates for low $l$ or low $\omega_{ln}$ (for fixed $n$). In order to perform an accurate integration in equation~\eqref{eqHij}, we have to use a very fine resolution in $\omega$ (see also Section \ref{secnumerics}). As a rough criterion, the peaks in the power spectrum with the largest contribution to total power should be well resolved. We thus aim to use a resolution which is smaller than the smallest linewidth when considering modes with a mode energy of $E_{nl}\propto\gamma_{nl} \Power(l,\omega_{nl}) \geq 10^{-3} \max_{l,n}(\gamma_{nl} \Power(l,\omega_{nl}))$, where $\Power$ is the unfiltered power spectrum and $l \leq l_\mathrm{max}=170$. This is the case when $\Delta \omega = 2 \pi / (\Delta t N_t)$ with $\Delta t=60\,\text{s}$ and $N_t=512000$. As a result, for modes with $\gamma_{nl} \Power(l,\omega_{nl}) \geq 10^{-2} \max(\gamma_{nl} \Power(l,\omega_{nl}))$, there are at least 12 grid points within the line width of a peak in the power spectrum and the most relevant part of the power spectrum is thus adequately covered.

In these considerations, a compromise had to be made between a good accuracy on the one hand and a reasonable computation time on the other hand. A characteristic computation with the above parameters using a range of modes with $l\leq 170$ and $0\leq n \leq 35$ takes roughly 2.5 days on 31 CPU cores. For a practical use in meridional flow inversions, where at least one kernel per travel distance has to be computed, an improvement concerning computation time seems desirable and remains subject to further research.

\begin{deluxetable}{ccccc}
\tabletypesize{\scriptsize}

\tablecolumns{5}
\tablecaption{Key Characteristics of $\Delta=42\degr$ Example Kernels\label{tablebigkernels}}
\tablewidth{0pt}
\tablehead{
\colhead{Kernel} & \colhead{Filter} &
\colhead{Mean $l$} & \colhead{Mean $\nu$ [mHz]} 
} 
\startdata
$K_1$ & $l\leq170$ & 84 & 2.933 \\ 
$K_2$ & $l\leq99$ & 49 & 2.929 \\  
$K_3$ & $l \leq 79$ & 39 & 2.928 \\ 
$K_4$ & $l \leq 49$ & 24 & 2.927 \\
$K_5$ & Gaussian ($l_0=45$, $\delta l=8$), $15\leq l\leq 75$  & 45 & 2.928 \\  
$K_6$ & phase-speed \citep{Kholikov2014}, $l\leq170$ & 46 & 2.995 
\enddata
\tablecomments{Mean $l$ and mean $\nu$ are computed as power-weighted averages.}
\end{deluxetable}

\subsection{Meridional Flow Kernels for Different Filters and $\Delta=42\degr$:  Results}
\label{seckernelresults}

The use of different filters results in a different spatial distribution of sensitivity in the kernels, see Figure~\ref{figbigkernels}. Kernel $K_{1}$ shows a pattern of large amplitude sensitivity with alternating sign close to the surface especially near the observation points. 
From kernel $K_{2}$ to $K_{4}$, where the higher-degree modes are successively filtered out, this ringing-like pattern disappears. 
The pattern shows a horizontal spatial scale which corresponds to $l\approx 260$. A multiplication of two Legendre polynomials as in equation~\eqref{eqZij} results in a summation of the corresponding degrees $l$. The pattern is thus produced by a range of modes of about $90\lesssim l \lesssim 170$. 
In addition to the kernels presented in Figure~\ref{figbigkernels}, we verified that a kernel computed using exclusively higher-degree modes ($140\leq l\leq 170$) shows a similar sensitivity pattern near the surface as does $K_1$. 
The ringing-like pattern may thus be attributed to the influence of higher-degree modes. 

In the lower-degree kernels presented in the bottom row of Figure~\ref{figbigkernels}, for which predominantly modes with $l \lesssim 70$ were used, the sensitivity is concentrated in deeper regions and the overall pattern is closer to a ``ray-path-like'' structure. 
It is noteworthy that the dominant negative sensitivity pattern present in 
$K_{5}$ at about $0.7-0.8R_\sun$ between the observation points in Figure~\ref{figbigkernels} is also visible in kernels $K_{1}$ to $K_{4}$. The sensitivity of the low degree modes to deep flows can thus be similarly present in deeper regions of kernels calculated with a larger set of modes, which illustrates the structure of the problem as a superposition of modes.

Furthermore, we find that the phase-speed filtered kernel $K_{6}$ is best localized with respect to depth. Although it shows the lowest amplitude, the integrated sensitivity has a magnitude comparable to the other kernels due to fewer locations with a sign reversal in the sensitivity. 
The general pattern of its sensitivity distribution in the spatial domain is found to be rather ``rectangular'' with distinct bands in radial direction. Such features have also been observed in Cartesian kernels (see bottom left panel in Fig. 1 from \citealp{BG2007}). Visually, the closest match to a ray-path-like pattern is produced by applying a Gaussian band-pass filter in $l$ ($K_5$) or by applying a simple cut-off filter in angular degree ($l\leq49$, $K_4$).

\section{SUMMARY AND DISCUSSION}

\label{secdiscussion}

In this paper, we have derived a formula for calculating the sensitivity of helioseismic travel times to flows with the Born approximation in spherical geometry. Following \citet{GB2002} and \citet{BG2007}, we solved the zero- and first-order problems in spherical geometry via Green's functions and expanded the kernel formula in a way such that it was implementable into numerical code.

The kernels can be used for inferring flows, which are small and have a linear effect on the travel times (e.g., \citealp{Jackiewicz2007a}) and where the influence of magnetic activity on the travel times can be neglected. 
In an example study for unfiltered data and a travel distance of 42 degrees, we found that uniform flows up to $500\,\rm{m}\,\rm{s}^{-1}$ can be treated with this approximation. For studying, e.g., meridional flow, the assumption of linearity is thus adequate as meridional flows are around $20\,\rm{m}\,\rm{s}^{-1}$ at the surface.

Our method was tested in two different ways. As a first sanity check, we considered an $f$-mode example for which the assumption of Cartesian geometry is appropriate. Results from our computation were compared to results from a Cartesian code used by \citet{BG2007}. 
We found a good qualitative and quantitative agreement between both results. Horizontal and total integrals, which correspond to the sensitivity of travel times to a uniform flow field, agree to within 0.3~\%. 
However, the amplitude of the Cartesian kernel was found to be more oscillatory compared to the kernel obtained in spherical geometry. 
This might be due to the approximations made in Cartesian geometry and the differences in the eigenfunction computation.

Secondly, as a test for self-consistency in the case of deep flows, we considered a travel distance of 42 degrees and uniform rotation. 
We found that the total integral of the kernel and the analytically-obtained reference value agree to within 2~\%. 
A high frequency resolution of $\Delta \omega = 2.045 \cdot 10^{-7}\,\mathrm{Hz}$, which is in the order of the lowest damping rates with a substantial contribution to the total power, is needed to well resolve the peaks in the power spectrum where the line widths are small. Due to this, the numerical implementation relied on a different approach compared to the one used by \citet{BG2007} for kernels in Cartesian geometry, where the power is concentrated on modes with higher $l$ and larger line widths.

Consequently, we used our method to evaluate the effect of different possible filters on the sensitivity of travel-time measurements to meridional flow for a travel distance of 42 degrees.
The sensitivity calculated using a wide range of modes ($l\leq170$) was found to be largely concentrated near the surface. A ringing-like pattern of large amplitude sensitivity with alternating sign was found especially between the observation points near the surface unlike in ray theory. This pattern is due to the inclusion of higher-degree modes (roughly $90\lesssim l \lesssim 170$). This finding is in accordance with the result of \citet{Bogdan1997} who showed that the wave path of a wave packet can be extended over a relatively large region from beneath the turning point of the corresponding ray path to the surface of the Sun.

When mainly lower-degree modes ($l \lesssim 70$) were used for obtaining filtered kernels, we found the sensitivity rather to be concentrated in regions near the bottom of the convection zone and rather to show a ``ray-path-like'' spatial distribution of the sensitivity. The ringing disappeared or it was significantly reduced. 
Among the filters presented here, it is noteworthy that a simple cut-off filter ($l\leq49$) and a narrow Gaussian filter in angular degree yielded sensitivity patterns which are visually most similar to a ray path. To our knowledge, a Gaussian filter in angular degree has not yet been used in helioseismic data analysis and may be worth exploring (see also \citealp{Jackiewicz2007a}). For phase-speed filtered measurements, kernels were found to be best localized in depth when compared to the examples presented in this paper.

As a further consequence for interpreting observations, we note that travel times have to be interpreted with caution (see also, e.g., \citealp{DeGrave2014QS} and \citealp{Svanda2015}). Different filters used in the data analysis process may lead to significant differences in the kernels.  
Similarly, we found for kernels computed in Cartesian geometry \citep{BG2007} that a number of changes in the parameters, which caused a change in the power-weighted mean frequency by about $0.2\;\rm{mHz}$, can cause the integrated sensitivity to change by a factor of 1.6 and extreme values to change by a factor of five. 
It is an open question, however, how inversion results are affected by such changes in the kernels.

\acknowledgments

Author contributions: A.C.B. and L.G. provided the theoretical derivation of the kernels in spherical geometry. A.C.B. also contributed numerical results for the Cartesian kernels in Section~\ref{secsanity}. V.B., M.R., and W.Z. have obtained the numerical results (Sections~\ref{secspecific} - \ref{secdeepkernels}). 
The authors thank Jesper Schou and Sylvain Korzennik for supplying damping rates. 
V.B. thanks Ariane Schad for commenting an earlier version of the manuscript, as well as Jason Jackiewicz, Shravan Hanasoge, Thomas Duvall, and
Junwei Zhao for valuable remarks and discussion. The research leading to these results has received funding from the European Research Council under the European Union’s Seventh Framework Program (FP/2007-2013)/ERC Grant Agreement no. 307117 (ORIGIN). 
L.G. and A.C.B. acknowledge support from DFG SFB 963 ``Astrophysical Flow Instabilites and Turbulence'' (Project A1). L.G, A.C.B., and M.R. acknowledge funding from EU FP7 Collaborative Project ``Exploitation of Space Data for Innovative Helio- and Asteroseismology'' (SpaceInn). L.G.  acknowledges support by the Center for Space Science at the NYU Abu Dhabi Institute under grant G1502.


\appendix

\section{FOURIER TRANSFORM CONVENTION}
\label{appendixfourier}

In the time domain, we adopt the Fourier transform convention of \cite{GB2002}. The same symbol is used for a function $f(t)$ and its Fourier transform $f(\omega)$ indicating the Fourier transform by using the frequency variable $\omega$. The Fourier transform is defined as
\begin{equation}
f(\omega) =  \frac{1}{2\pi} \int_{-\infty}^\infty f(t) \, e^{i\omega t} \, \id t.
\end{equation}
For this convention, Parseval's theorem reads
\begin{equation}
                \int_{-\infty}^\infty f(t) \, g^*(t) \, \id t = 2\pi \int_{-\infty}^\infty  f(\omega) \,g^*(\omega) \, \id \omega .
                \label{eqparseval}
\end{equation}

\section{ZERO ORDER SOLUTION AND GREEN'S FUNCTIONS}
\label{appendixgreens}

We show first that the ansatz \eqref{eqansatz} solves the wave equation~\eqref{eqLfreqspace}, provided the Green's functions in turn satisfy equation~\eqref{eqgreens}. 
We recall that the operator $\cL$ acts only on the $\br$-dependence of $\bG^k(\br|\br',\omega)$ taking $\br'$ and $\omega$ to be constant. Assuming \eqref{eqgreens} to be fulfilled, we plug our ansatz \eqref{eqansatz} into the zero-order wave equation~\eqref{eqLfreqspace} and calculate
\begin{align}
        \cL [\xi_j(\br,\omega)] &= \int_\sun   \cL \Big[ G^k_j(\br|\br',\omega) S_k(\br',\omega) \Big] \, \id^3 \br' \\
        &=  \int_\sun  \unit_j^{(k)}(\br') \delta(\br-\br')  S_k(\br',\omega) \,\id^3 \br' = S_j(\br,\omega),
\end{align}
which shows that indeed, the ansatz \eqref{eqansatz} solves the wave equation, provided equation~\eqref{eqgreens} holds. In order to solve equation~\eqref{eqgreens} to obtain an expression for the Green's function, we now expand $\bG$ on the eigenfunction basis,
\begin{equation}
\bG^k(\br|\br',\omega) = \sum_{lmn} c_{lmn}^k(\br',\omega)\bxi^{lmn}(\br) .
\label{eqgreensexpansion}
\end{equation}
We start with the right-hand side of \eqref{eqgreens} and use \eqref{eqgreensexpansion} and \eqref{eqdamping} on the left-hand side of \eqref{eqgreens}, which yields for $k=r,\theta,\phi$,
\begin{equation}
 \unit^{(k)}(\br') \delta(\br-\br')=\cL [\bG^k(\br|\br',\omega) ]= \sum_{l'm'n'} c_{l'm'n'}^k(\br',\omega) (-\omega^2 + \omega_{l'n'}^2 -2\ii\omega\gamma_{l'n'}) \rho_0(\br) \bxi^{l'm'n'}(\br) .
\end{equation}
Multiplying both sides of this equation with $\bxi^{lmn*}(\br)$ and integrating over the whole volume of the Solar model in the $\br$ domain, we obtain with \eqref{eqorthonormal},
\begin{align}
\xi_k^{lmn*} (\br') 
&= \sum_{l'm'n'} c_{l'm'n'}^k(\br',\omega) (-\omega^2 + \omega_{l'n'}^2 -2\ii\omega\gamma_{l'n'}) \int_\sun \rho_0(\br) \bxi^{lmn*}(\br)\bxi^{l'm'n'}(\br) \, \id^3 \br  \\
&= c_{lmn}^k(\br',\omega) (-\omega^2 + \omega_{ln}^2 -2\ii\omega\gamma_{ln}) ,
\end{align}
from which we find
\begin{align}
c_{lmn}^k(\br',\omega) = \frac{ \xi_k^{lmn*} (\br') }{-\omega^2 + \omega_{ln}^2 -2\ii\omega\gamma_{ln}} .
\end{align}
From \eqref{eqgreensexpansion}, we obtain for the Green's function,
\begin{equation}
G_j^k(\br|\br',\omega)  = \sum_{lmn} \frac{ \xi_k^{lmn*} (\br') \xi_j^{lmn} (\br) }{\sigma^2_{ln}-\omega^2 } 
 \label{eqgreensformula}
\end{equation}
and
\begin{equation}
G_r^k(\br|\br',\omega)  = \sum_{lm} \left(\sum_n \frac{  \xi_k^{lmn*} (\br')  }{\sigma^2_{ln}-\omega^2 } R_{ln}(\robs) \right) Y_{lm} (\theta,\phi), \label{eqgreensforfiltering}
\end{equation}
where
\begin{equation}
\sigma^2_{ln} = \omega_{ln}^2 -2\ii\omega\gamma_{ln}.
\end{equation}
In the time domain, the Green's functions
\begin{equation}
\hat G_j^k(\br,\br',t,t')=\int_{-\infty}^\infty G_j^k(\br|\br',\omega)\, e^{-\ii\omega(t-t')} \, \id\omega
\label{eqgreenstimedomain}
\end{equation}
thus satisfy (see also equation~[9] in \citealp{BG2007})
\begin{equation}
\cL[\hat \bG^k(\br,\br',t,t')]=2\pi \, \unit^{(k)}(\br') \, \delta(\br-\br') \, \delta(t-t').
\label{eqgreenstimedomaincondition}
\end{equation}

In addition to the Green's functions fulfilling equation~\eqref{eqgreens}, or equivalently equation~\eqref{eqgreenstimedomaincondition}, any zero-order solution of the wave equation needs to fulfil spatial and temporal boundary conditions. For a detailed treatment of the spatial boundary conditions for the eigenfunctions at the center of the Sun and at the outer boundary, we refer to \citet{JCDadipls}, \citet[section 3.3.2.2]{Aerts2010}, and to \citet[chapters 14 and 18]{Unno1989}. 
The spatial boundary conditions are satisfied by the Green's functions $G_j^k$ and by any zero-order solution $\bxi$ obtained with equation~\eqref{eqansatz} due to the linearity of the problem if the eigenmodes fulfil the same spatial boundary conditions. In addition, the condition $\bxi(\br,t)=0$ needs to be satisfied for $t\leq-T/2$ due to equation \eqref{eqzeroorder} as $\bS(\br,t)=0$ for $t\leq -T/2$. This is guaranteed by equation \eqref{eqansatz} and the Green's functions, which satisfy $G_j^k(\br|\br',t-t')=0$ for $t-t'\leq 0$.

\section{ZERO-ORDER POWER SPECTRUM}
\label{appendixpower}

Filtered spherical harmonic coefficients $a_{lm}(\omega)$ defined in equation~\eqref{eqalmfiltered2} can be obtained from equations \eqref{eqSHT}, \eqref{equnfilteredsignal}, \eqref{eqansatz}, 
and \eqref{eqgreensforfiltering}. We find
\begin{equation}
a_{lm}(\omega) = -\ii\omega f(l,\omega) \sum_{n} \frac{R_{ln}(\robs)}{\sigma^2_{ln}-\omega^2 } \int_\sun  \xi_k^{lmn*} (\br') S_k(\br',\omega) \, \id^3 \br',  \label{eqalmw}
\end{equation}
from which we deduce the zero order power spectrum
\begin{align}
        \Power_0(l,\omega) &= \EE[\Power(l,\omega)] = \frac{2\pi}{T}\, \EE[ \sum_{m=-l}^l  |a_{lm}(\omega)|^2 ] = \frac{2\pi}{T}\, \sum_{m=-l}^l \EE [a_{lm}^*(\omega)a_{lm}(\omega) ]  \\
        & = \frac{2\pi}{T}\, \omega^2 f(l,\omega)^2 \sum_{mnn'} R_{ln}(\robs) R_{ln'}(\robs) \nonumber \\ 
				& \quad \times\,\EE [ \Big( \int_\sun  \frac{  \xi_j^{lmn*} (\br') }{\sigma^2_{ln}-\omega^2 } S_j(\br',\omega) \, \id^3 \br'\Big)^* \Big( \int_\sun  \frac{  \xi_k^{lmn'*} (\br'') }{\sigma^2_{ln'}-\omega^2 } S_k(\br'',\omega) \, \id^3 \br''\Big) ] \label{eqlinebreak1}  \\
        &=  \omega^2 f(l,\omega)^2 \sum_{mnn'} \frac{R_{ln}(\robs) R_{ln'}(\robs)}{(\sigma^{2*}_{ln}-\omega^2)(\sigma^{2}_{ln'}-\omega^2)} \nonumber \\
				& \quad \times \,  \int_\sun \int_\sun  \xi_j^{lmn} (\br')  \xi_k^{lmn'*} (\br'') \, \frac{2\pi}{T}\, \EE [  S_j^*(\br',\omega)  S_k(\br'',\omega) ]  \, \id^3 \br' \id^3 \br''. \label{eqlinebreak2}\\
	&= \omega^2 M(\omega)  f(l,\omega)^2 \sum_{m=-l}^l \sum_{nn'} \frac{ R_{ln}(\robs) R_{ln'}(\robs) R_{ln}(r_s)R_{ln'}(r_s) }{(\sigma^{2*}_{ln}-\omega^2)(\sigma^{2}_{ln'}-\omega^2)},  \label{eqpowerusesourcecovariance}
\end{align}
\notetoeditor{The line breaks in equations \eqref{eqlinebreak1} and \eqref{eqlinebreak2} were introduced for readability in manuscript format and may be omitted.}which yields equation~\eqref{eq0power} when performing the sum over $m$. 
In equation~\eqref{eqpowerusesourcecovariance}, we used the relation 
\begin{align}
 &\int_\sun \int_\sun  \xi_j^{lmn} (\br')  \xi_k^{l'm'n'*} (\br'') \, \frac{2\pi}{T}\, \EE [  S_j^*(\br',\omega)  S_k(\br'',\omega) ]  \, \id^3 \br' \id^3 \br'' \\
&=M(\omega) \int_{S^2}  \frac{r_s^2}{r_s^2}  R_{ln}(r_s) R_{l'n'}(r_s) Y_{lm}(\Omega')  Y_{l'm'}^*(\Omega') \, \id \Omega' \label{eqintegraloversourcecovariancefirststep}\\
&=M(\omega) R_{ln}(r_s)R_{ln'}(r_s) \delta_{ll'} \delta_{mm'}, \label{eqintegraloversourcecovariance}
\end{align}
which was deduced using the properties of the source covariance defined in equation~\eqref{eqsourcecovariance} as well as $\id^3 \br'=r'^2 \id r' \id \Omega'$. Equation~\eqref{eqintegraloversourcecovariance} will also be useful in the following derivations.

\section{ZERO-ORDER CROSS-COVARIANCE}
\label{appendixcrosscorr}

Using our convention for the cross-covariance function from equation~\eqref{eqconvcorrw} and the expression for the filtered observational Doppler signal from equation~\eqref{eqdopplerfilteredgreens}, we obtain
\begin{align}
C_0(\br_1,\br_2,\omega) 
				&= \EE \left[ C(\br_1,\br_2,\omega) \right]
				=\frac{2 \pi}{T} \EE \left[ \Phi^*(\br_1,\omega)  \Phi(\br_2,\omega) \right] \label{eqcrosscorrw3} \\
        &= \frac{2\pi}{T} \EE \Big[ \; \left( -\ii\omega   \sum_{lmn} f(l,\omega)  \frac{ \xi_r^{lmn}(\br_1) }{\sigma^2_{ln}-\omega^2 }   \int_\sun  \xi_i^{lmn*} (\br') S_i(\br',\omega) \, \id^3 \br'  \right)^*\nonumber \\
        & \quad \times \left( -\ii\omega   \sum_{l'm'n'} f(l',\omega)  \frac{ \xi_r^{l'm'n'}(\br_2) }{\sigma^2_{l'n'}-\omega^2 }   \int_\sun  \xi_j^{l'm'n'*} (\br'') S_j(\br'',\omega) \, \id^3 \br''  \right) \; \Big]  \\
        &= \omega^2   \sum_{ll'mm'nn'} f(l,\omega)f(l',\omega) \frac{ \xi_r^{lmn*} (\br_1) }{\sigma^{2*}_{ln}-\omega^2 }  \frac{   \xi_r^{l'm'n'} (\br_2) }{\sigma^2_{l'n'}-\omega^2 } \nonumber  \\
        & \quad \times \int_\sun \int_\sun  \xi_i^{lmn} (\br') \xi_j^{l'm'n'*} (\br'') \, \frac{2\pi}{T}\,\EE \left[ \; S_i^*(\br',\omega)  S_j(\br'',\omega) \right] \, \id^3 \br' \, \id^3 \br''  \\
        &= \omega^2 M(\omega)  \sum_{lnn'} f(l,\omega)^2 \frac{ R_{ln}(r_s) }{\sigma^{2*}_{ln}-\omega^2 }  \frac{ R_{ln'}(r_s)  }{\sigma^2_{ln'}-\omega^2 }  \,  \sum_m  \xi_r^{lmn*} (\br_1)   \xi_r^{lmn'} (\br_2) \label{eqcrosscorrtrick} \\
        &=  \frac{\omega^2}{4\pi}  M(\omega)  \sum_{lnn'} (2l+1) f(l,\omega)^2 R_{ln}(r_s) R_{ln'}(r_s)   \frac{ R_{ln}(r_1) R_{ln'}(r_2) P_l(\cos \Delta_{1,2}) }{(\sigma^{2*}_{ln}-\omega^2) (\sigma^2_{ln'}-\omega^2 )}. 		\label{eq0crosscorrappendix}
\end{align}
In \eqref{eqcrosscorrtrick}, we used equation~\eqref{eqintegraloversourcecovariance}. In equation~\eqref{eq0crosscorrappendix}, we applied
\begin{align}
 \sum_m  \xi_r^{lmn*} (\br_1)   \xi_r^{lmn'} (\br_2)&= R_{ln}(r_1) R_{ln'}(r_2) \sum_m Y_{lm}^*(\Omega_1)Y_{lm}(\Omega_2) \\
& = R_{ln}(r_1) R_{ln'}(r_2) \frac{2l+1}{4\pi}P_l(\cos \Delta_{1,2}),
\end{align}
where $\Delta_{1,2}=\Delta(\Omega_1,\Omega_2)$ denotes the angular distance between positions $\Omega_1$ and $\Omega_2$ on the unit sphere, and $P_l$ denotes a Legendre polynomial of degree $l$.

\section{GENERAL KERNEL FORMULA: DERIVATION}
\label{appendixgeneral}

For deriving the general kernel formula, we use equation~\eqref{eqdeltaxi} for the first-order perturbed wave field and the expression for the zero order wave field from equation~\eqref{eqansatz},
\begin{align}
        \delta \bxi_j(\br_2,\omega)   &=  2\ii\omega  \int_\sun \int_\sun  \rho_0(r) G^k_j(\br_2|\br,\omega) \, \bv(\br)   \bcdot \bnabla_{\br} [  G^i_k(\br|\br',\omega) S_i(\br',\omega) ] \, \id^3 \br \, \id^3 \br'   , 
\end{align}
from which we obtain for the filtered Doppler signal with equation~\eqref{eqdeltaPhi},
\begin{align}
   \delta \Phi(\br_2,\omega)   &= 2 \omega^2   \int_\sun \int_\sun  \rho_0(r)  {\cG}^k(\br_2|\br,\omega) \, \bv(\br)   \bcdot \bnabla_{\br} [ G^h_k(\br|\br',\omega) S_h(\br',\omega) ] \, \id^3 \br \, \id^3 \br' \label{eqdeltaPhi2}.
\end{align}
We can now deduce an expression for $\delta C$, neglecting the second-order term and using equation~\eqref{eqdopplerfilteredgreens},
\begin{align}
        \delta C(\br_1,\br_2,\omega) &= \frac{2 \pi}{T} \EE \Big[ \Phi^*(\br_1,\omega)  \delta \Phi(\br_2,\omega) + \delta \Phi^*(\br_1,\omega)  \Phi(\br_2,\omega) \Big] \\
        &= \frac{2 \pi}{T}\EE \Big[ \ii \omega  \int_\sun  {\cG}^{j*}(\br_1|\br'',\omega)  S_j^*(\br'',\omega) \, \id^3 \br''  \nonumber \\
        & \;\;\; \times 2 \omega^2   \int_\sun \int_\sun  \rho_0(r)  {\cG}^{k}(\br_2|\br,\omega) \bv(\br)   \bcdot  \bnabla_{\br} [  G^{h}_k(\br|\br',\omega)] S_h(\br',\omega) \, \id^3 \br \, \id^3 \br' \Big] + (1 \leftrightarrow 2)^*  \\
        &= \int_\sun \bv(\br)   \bcdot \bcurlyC(\br_1,\br_2,\omega;\br) \, \id^3 \br ,
\end{align}
where we indicate with $(1 \leftrightarrow 2)^*$ an additional term which is identical to the previous one except for complex conjugation and exchange of indices 1 and 2, and where we obtain
\begin{align}
  \bcurlyC(\br_1,\br_2,\omega;\br)  
        &= 2 \ii\, M(\omega) \, \omega^3   \rho_0(r) \, {\cG}^{k}(\br_2|\br,\omega) \int_{S^2}  \bnabla_{\br} [  G^{r}_k(\br|r_s, \Omega',\omega)] \, {\cG}^{r*}(\br_1|r_s, \Omega',\omega) \, \id  \Omega'  + (1 \leftrightarrow 2)^* \label{eqc1}.
\end{align}
Equation~\eqref{eqc1} is deduced using the expression for the source covariance from equation~\eqref{eqsourcecovariance} proceeding similarly as in equation~\eqref{eqintegraloversourcecovariancefirststep} and indicating a particular source location with $\br_s=(r_s,\Omega')$. Equation~\eqref{eqc1} is used in Section \ref{secgeneral} to derive the general kernel formula \eqref{eqkernelgeneral}.

\section{SPECIFIC KERNEL FORMULA: DERIVATION}
\label{appendixspecific}

We first evaluate some terms in equation~\eqref{eqkernelgeneral} using the expression for the filtered Green's functions in equation~\eqref{eqfilteredgreens}. Keeping in mind $\cO_k^{ln*} = \cO_k^{ln}$, we find with $\br=(r,\Omega)$ and $\br_2=(r_2,\Omega_2)$,
\begin{align}
        {\cG}^{k}(\br_2|\br,\omega)
				&= \sum_{\bar l \bar m \bar n} f(\bar l,\omega) \frac{  \xi_k^{\bar l \bar m \bar n *} (\br) \xi_r^{\bar l \bar m \bar n} (\br_2) }{\sigma^{2}_{\bar l \bar n}-\omega^2 }  \\
        &= \sum_{\bar l \bar n} f(\bar l,\omega) \frac{ R_{\bar l \bar n}(r_2) }{\sigma^{2}_{\bar l \bar n}-\omega^2 }  \cO_k^{\bar l \bar n*} (\br)\big[ \sum_{\bar m} Y_{\bar l \bar m }^* (\Omega) Y_{\bar l \bar m }(\Omega_2) \big]  \\
        &= \sum_{\bar l \bar n} f(\bar l,\omega) \frac{2\bar l +1}{4\pi}\frac{ R_{\bar l \bar n}(r_2) }{\sigma^{2}_{\bar l \bar n}-\omega^2 }  \cO_k^{\bar l \bar n} (\br)\big[ P_{\bar l}(\cos \Delta_2) \big] \label{eqGtilde},
\end{align}
where $\Delta_2=\Delta(\Omega,\Omega_2)$. Next, taking into account that both $\bnabla_{\br}$ and $\cO_k^{ln}(\br)$ act on an $\br$-dependence, we obtain from equations \eqref{eqfilteredgreens} and \eqref{eqgreensformula},
\begin{align}
         \int_{S^2}  \bnabla_{\br} &[  G^{r}_k(\br|r_s, \Omega',\omega)] \, {\cG}^{r*}(\br_1|r_s, \Omega',\omega) \, \id  \Omega'\label{eqconvolutionGG} \\
        =&  \sum_{ll'mm'nn'} \int_{S^2}  \bnabla_{\br} \left[ \frac{  \xi_r^{lmn*} (r_s,\Omega') \xi_k^{lmn} (\br) }{\sigma^{2}_{ln}-\omega^2 }\right] f(l',\omega) \frac{  \xi_r^{l'm'n'} (r_s, \Omega')  {\xi}_r^{l'm'n'*} (\br_1) }{\sigma^{2*}_{l'n'}-\omega^2 } \, \id  \Omega'  \\
        =&  \sum_{ll'mm'nn'} \int_{S^2}  Y_{lm}(\Omega') Y_{l'm'}^*(\Omega') \id  \Omega' \, f(l',\omega) R_{ln}(r_s)R_{l'n'}(r_s)   \frac{  \bnabla_{\br} \left[ \xi_k^{lmn} (\br) \right]  {\xi}_r^{l'm'n'*} (\br_1) }{(\sigma^{2}_{ln}-\omega^2) (\sigma^{2*}_{l'n'}-\omega^2) }  \\
        =& \sum_{lnn'}   f(l,\omega) \frac{ R_{ln}(r_s)R_{ln'}(r_s)  }{(\sigma^{2}_{ln}-\omega^2) (\sigma^{2*}_{ln'}-\omega^2) } \sum_m  \bnabla_{\br} \left[ \xi_k^{lmn} (\br) \right]  \xi_r^{lmn'*} (\br_1) \\
				=& \sum_{lnn'}  f(l,\omega)  \frac{ R_{ln}(r_s)R_{ln'}(r_s) R_{ln'}(r_1) }{(\sigma^{2}_{ln}-\omega^2) (\sigma^{2*}_{ln'}-\omega^2) }  \bnabla_{\br} \Bigg[   \cO_k^{ln} (\br) \Big[ \sum_m Y_{lm}(\Omega)Y_{lm}^*(\Omega_1)\Big]     \Bigg] \\
        =& \sum_{lnn'}  f(l,\omega) \frac{2l+1}{4\pi}  \frac{ R_{ln}(r_s)R_{ln'}(r_s) R_{ln'}(r_1) }{(\sigma^{2}_{ln}-\omega^2) (\sigma^{2*}_{ln'}-\omega^2) }
 \bnabla_{\br} \Bigg[   \cO_k^{ln} (\br) \Big[  P_l(\cos \Delta_1)  \Big]     \Bigg]\label{eqintnablaGG}.
\end{align}
Plugging now equations \eqref{eqGtilde} and \eqref{eqintnablaGG} into the general kernel formula \eqref{eqkernelgeneral} and rearranging terms, we obtain
the desired spherical kernel formula \eqref{eqcalckernel}.


\bibliographystyle{apj}

\providecommand{\noopsort}[1]{}\providecommand{\singleletter}[1]{#1}%


\end{document}